\documentclass[11pt,a4paper]{article}

\usepackage[utf8]{inputenc}
\usepackage{bold-extra}
\usepackage{chngcntr}
\usepackage{tikz}
\usepackage{setspace}
\usepackage{url}
\usepackage{amsfonts}
\usepackage{amssymb}
\usepackage{epsfig}
\usepackage{amsmath}
\usepackage{commath}
\usepackage{natbib}
\usepackage{graphicx}
\usepackage{bbm,float}
\usepackage{dsfont}
\usepackage{color}
\usepackage{adjustbox}
\usepackage{amsmath,amssymb,amstext,amsfonts,amsthm,latexsym}
\usepackage{diagbox}
\usepackage{bm}
\usepackage{caption}
\usepackage{subcaption}
\usepackage{multirow}
\usepackage{color}
\usepackage{listings}
\usepackage{titlesec}
\usepackage{enumerate}
\usepackage{environ}
\usepackage{booktabs}
\usepackage[linesnumbered,ruled]{algorithm2e}
\usepackage{appendix}
\usepackage{bbm}
\usepackage{pgfplots}
\usepackage{makecell}
\usepackage{tabularx}
\newcolumntype{Y}{>{\centering\arraybackslash}X}
\numberwithin{equation}{section}
\theoremstyle{plain}
\newtheorem{theorem}{Theorem}[section]
\newtheorem{corollary}{Corollary}[section]
\newtheorem{proposition}{Proposition}[section]

\newtheorem{lemma}{Lemma}[section]

\theoremstyle{definition}

\newtheorem{definition}{Definition}[section]

\theoremstyle{remark}

\newtheorem{remark}{Remark}[section]
\pgfplotsset{compat=1.17}

\begin{document}
\title{\textbf{Higher moments under dependence uncertainty with applications in insurance}
}
\date{\today}
\author{Carole Bernard\thanks{Carole Bernard, Department of Accounting, Law and Finance, Grenoble Ecole de Management (GEM) and Department of Economics at Vrije Universiteit Brussel (VUB).  (email: \texttt{carole.bernard@grenoble-em.com}).}, Jinghui Chen\thanks{Corresponding author: Jinghui Chen, Department of Mathematics and Statistics at York University, and RISC Foundation.  (email: \texttt{jh8chen@yorku.ca}).}, and Steven Vanduffel\thanks{Steven Vanduffel, Department of Economics at Vrije Universiteit Brussel (VUB). (e-mail: \texttt{steven.vanduffel@vub.be}). }}
\maketitle
\begin{abstract}
		Recent studies have highlighted the significance of higher-order moments - such as coskewness - in portfolio optimization within the financial domain. This paper extends that focus to the field of actuarial science by examining the impact of these moments on key actuarial applications. In the first part, we derive analytical lower and upper bounds for mixed moments of the form $\mathbb{E}(X_1X_2^d)$, where $X_i \sim F_i$ for $i=1,2$, assuming known marginal distributions but unspecified dependence structure. The results are general and applicable to arbitrary marginals and positive integer orders $d$, and we also identify the dependence structures that attain these bounds. These findings are then applied to bound centered mixed moments and explore their mathematical properties.
		
		The second part of the paper investigates the influence of higher-order centered mixed moments on key actuarial quantities, including expected shortfall (ES), marginal expected shortfall (MES), and life annuity valuation. Under a copula-based mixture model, we show that coskewness and other odd-order mixed moments exhibit a monotonic relationship with both ES and annuity premiums. However, the effect on MES is more nuanced and may remain invariant depending on the underlying dependence structure.
\end{abstract}

\textbf{Keywords:} Higher-order moments, Copula, Risk bounds, Coskewness, Expected shortfall, Marginal expected shortfall, Life annuity.

\newpage
\section{Introduction}
The role of higher-order moments - beyond variance - has been extensively investigated in economics, particularly in portfolio analysis grounded in utility theory \citep[see, e.g.,][]{samuelson1970fundamental, jondeau2006optimal, harvey2010portfolio, wang2025stochastic}. This line of inquiry was notably accelerated by widespread theoretical and empirical critiques of Markowitz modern portfolio theory \citep{scott1980direction}. 

In portfolio analysis, the expected utility of a portfolio's return $X$ may be approximated via an infinite order Taylor series expansion, that is
\begin{equation*}
	\mathbb{E}(\mu(X))=\mathbb{E}\left[ \sum_{i=0}^{\infty} \frac{\mu^{(i)}(\mathbb{E}(X)) (X - \mathbb{E}(X))^i}{i!} \right]
	= \sum_{i=0}^{\infty} \frac{\mu^{(i)}(\mathbb{E}(X))}{i!} \mathbb{E}\left[(X - \mathbb{E}(X))^i\right],
\end{equation*}where $\mathbb{E}(\cdot)$ denotes the expectation operator, $0!=1$, $\mu$ is a utility function and $\mu^{(i)}$ denotes its $i$-th derivative with $\mu^{(0)}(\cdot)=\mu(\cdot)$. The convergence of this expansion depends on the specific utility function used as well as the distribution of $X$ \citep{loistl1976erroneous, lhabitant1998ab}. \citet{fahrenwaldt2020expected} assess the accuracy of this expansion for utility-based optimization.

Let $\bm{X} = (X_1, X_2, \dots, X_d)'$ denote the vector of returns from $d$ risky assets, and let $\bm{\alpha} = (\alpha_1, \alpha_2, \dots, \alpha_d)'$ represent the portfolio weights. The overall return is then $X = \bm{\alpha}' \bm{X}$, and the expected utility becomes
\begin{equation*}
	\mathbb{E}(\mu(X))
	= \mu(\mathbb{E}(X))+\sum_{i=2}^{\infty} \frac{\mu^{(i)}(\mathbb{E}(X))}{i!} \sum_{j=1}^{d}\alpha_j\mathbb{E}\left[(X_j-\mathbb{E}(X_j))(X - \mathbb{E}(X))^{i-1}\right],
\end{equation*}
where the inner expectation denotes the $i$-th-order centered mixed moments between asset $X_j$ and the portfolio return $X$. For $i=1, 2, 3$, these moments correspond to covariance, non-standardized coskewness and cokurtosis, respectively.

These centered mixed moments are used in portfolio construction and asset pricing. Classical models such as the mean–variance efficient frontier \citep{markowitz1952portfolio} and the capital asset pricing model (CAPM) \citep{sharpe1964capital} rely on covariance to describe systematic risk through the beta coefficient. However, more recent studies have turned their attention to coskewness. For instance, \citet{harvey2000conditional} show that including coskewness in asset pricing models helps explain the cross-sectional variation in expected returns, particularly where traditional models fail. Likewise, \citet{kraus1976skewness} and \citet{friend1980co} introduce skewness-adjusted versions of CAPM.

Applications of coskewness span various financial domains including mutual funds, options, and stochastic control \citep{vanden2006option, smith2007conditional, moreno2009value, harvey2010portfolio, wang2025stochastic}. \citet{arbia2020least} provide a comprehensive overview of coskewness in financial risk measurement. Empirical evidence from \citet{bressan2022time} shows that during periods of market distress - such as the global financial crisis or the COVID-19 pandemic - bank returns and their coskewness with the market become positively correlated, in contrast to negative correlation during stable periods. \citet{dong2022co} further demonstrate that coskewness with market indices at different scales (local, regional, global) has a significant negative effect on expected returns. For cokurtosis, we refer the reader to \citet{jondeau2006optimal}, \citet{boudt2015higher}, \citet{vo2021higher}, \citet{le2024efficient}, and references therein.

Despite its importance in finance, the role of higher-order centered mixed moments in actuarial science remains underexplored. To the best of our knowledge, most actuarial research focuses primarily on second-order moments. For example, \citet{furman2008weighted} reformulate weighted risk capital allocations as a function of covariance. \citet{godin2019general} apply second-order pricing principles in insurance, and \citet{furman2009weighted} introduce the weighted insurance pricing model (WIPM) as an actuarial analogue to CAPM. \citet{furman2017beyond} extend WIPM to Gini-type frameworks, accommodating settings where second-order moments may be infinite.

This paper seeks to address this gap by investigating how higher-order centered mixed moments of the form$$\mathbb{E}[(X_j - \mathbb{E}(X_j))(X - \mathbb{E}(X))^{i-1}], \quad i > 2,$$affect three core actuarial areas: insurance capital requirements, risk capital allocations, and life annuity valuation. For the first two, we adopt ES and MES as risk measures due to their well-established theoretical foundations. ES is known to be coherent \citep{artzner1999coherent, acerbi2002expected}, comonotonically additive \citep{kusuoka2001law}, convex \citep{follmer2002convex}, and elicitable \citep{fissler2016higher}. It is uniquely characterized by economic axioms \citep{wang2021axiomatic}, and is categorized as both a distorted \citep{wang2000distortion} and weighted \citep{furman2008weighted} risk measure. The regulatory Solvency III framework, consistent with Basel III's Fundamental Review of the Trading Book \citep{basel2016es}, plans to replace Value-at-Risk (VaR) at $p = 0.99$ with ES at $p = 0.975$ for market risk capital assessments. MES, as a derivative of ES, inherits many of these favorable properties \citep[e.g.,][and references therein]{tsanakas2003risk, furman2008weighted, dhaene2012optimal}. For the annuity setting, we consider joint-life and last-survivor annuities, as described by \citet{denuit2001measuring} and \citet{dickson2020actuarial}, which represent pension guarantees for couples.

In the first part of this paper, we derive analytical lower and upper bounds on mixed moments of the form $\mathbb{E}(X_1 X_2^d)$, for known marginals $F_1\sim X_1$ and $F_2\sim X_2$ but unknown dependence. These results extend the work of \citet{bernard2023coskewness} on mixed moments of the form $\mathbb{E}(X_1 X_2 \dots X_d)$. We then apply these bounds to study the centered mixed moments and propose novel dependence measures: standardized rank coefficients, taking values in $[-1, 1]$. This complements the earlier, non-standardized rank coskewness by \citet{bucher2017testing} and is in the same spirit as Spearman's rank correlation \citep{spearman1961proof}.

The second part of the paper explores how higher-order standardized centered mixed moments influence ES, MES, and life annuity pricing. To facilitate this, we assume exponential marginal distributions and employ a mixture copula that blends the extremal copulas maximizing and minimizing the mixed moments. This choice of exponential marginals is justified by their widespread use in modeling mortality and insurance losses \citep{dickson2020actuarial}, as well as their analytical tractability. For practical calibration, we use real demographic data from Statistics Canada \citep{statcan_life_expectancy, statcan_ave_re}.
 
This study offers several critical contributions: First, the lower and upper bounds on any order mixed moments are derived and the explicit dependence structures for arbitrary marginal distributions to reach them are obtained. These bounds are crucial for stress testing under extreme dependence scenarios, such as financial crises or pandemics. Second, we introduce the standardized rank coefficients, robust and distribution-invariant dependence measures that enable consistent comparisons across datasets.

Beyond these theoretical results, our conclusions have meaningful implications for practitioners. For instance, insurers aiming to lower capital requirements under regulatory frameworks may consider product lines with lower odd-order mixed moments relative to their existing portfolios. However, caution is warranted: reducing coskewness at the asset level does not necessarily lower systemic risk contribution, as MES may remain unaffected or even increase depending on the parameters and marginal distributions. In the realm of life annuities, higher-order mixed moments - especially odd-order terms capturing dependency between future lifetimes - can have a substantial impact on pricing. Assuming incorrect dependence models may lead to mispricing.

The paper is organized as follows. Section~\ref{risk bounds} derives upper and lower bounds on mixed moments of the form $\mathbb{E}(X_1 X_2^d)$, along with associated dependence structures. We apply these results to the study of coskewness and introduces the standardized rank coskewness in Section~\ref{first properties of coskewness}. Section~\ref{po ap} analyzes the effect of higher-order moments on ES, MES, and annuity pricing under the proposed mixture model. The last section draws our conclusion. 

\section{Sharp lower and upper bounds}\label{risk bounds}
Throughout this paper, let \((\Omega, \mathcal{A}, \mathbb{P})\) denote a standard atomless probability space, and let $L^0 := L^0(\Omega, \mathcal{A}, \mathbb{P})$ represent the set of all real-valued random variables $X,\ Y,\ X_i,\ L_i,\ i\in\{1,2\},\ T_x,\ T_y$ on this probability space. $d,\ k \in \mathbb{N}^+$ denote two positive integer numbers. The notation $X\sim F_X$ indicates that the random variable $X$ follows the cumulative distribution function (CDF) \(F_X(x) = \mathbb{P}(X \leq x), \, x \in \mathbb{R}\). The function \(F_i^{-1}(p)\) for $p \in (0, 1)$ denotes the generalized quantile function of \(F_X\) defined as  
\[
F_X^{-1}(p) = \inf \{x \in \mathbb{R} : F_X(x) \geq p \}.
\] Unless otherwise specified, we assume that all random variables involved have finite expectations, and $U$ denotes a standard uniform random variable.

The central question of this section is to derive lower and upper bounds on the expectation of the product of $X_1$ and $X_2^d$ with known marginal distributions but unknown dependence structure. Specifically, we consider the problems
\begin{subequations}
	\label{sharp bounds}
	\begin{align}
		m&=\inf \limits_{\forall i\ X_i\sim F_i, i=1,2}\mathbb{E}\left(X_1X_2^d\right); \label{m} \\
		M&=\sup \limits_{\forall i\ X_i\sim F_i, i=1,2}\mathbb{E}\left(X_1X_2^d\right).\label{M}
	\end{align}
\end{subequations}
There exist some similar but different problems in the literature. For example, \cite{wang2011complete} obtain a lower bound on the expectation of the product of $n>2$ random variables, when their distributions are completely mixable. Moreover, under various assumptions on the marginal distributions, \cite{bernard2023coskewness} derive the bounds on the expectation of the product of $n>2$ random variables, i.e., $\mathbb{E}(X_1X_2\dots X_n)$. However, $m$ in \eqref{m} and $M$ in \eqref{M} for $d\geq2$ have not been studied as far as we know.

\subsection{Analytic results}
In this subsection, we solve Problems~\eqref{m} and \eqref{M} analytically. We first recall the following two well-known results on the bounds of the expectation of the product of two variables.

\begin{lemma}[Maximum Product] \label{max product}
	Let $X\sim F$, $ Y\sim G$ and $U\sim U[0,1]$. Then,
	\begin{equation}\label{max ineq}
		\sup \limits_{X\sim F,Y\sim G}\mathbb{E}(XY)=\mathbb{E}\left(F^{-1}(U)G^{-1}(U)\right).
	\end{equation}
\end{lemma}

\begin{lemma}[Minimum Product] \label{min product}
	Let $X\sim F$, $ Y\sim G$ and $U\sim U[0,1]$. Then, \begin{equation}\label{min ineq}
		\inf \limits_{X\sim F,Y\sim G}\mathbb{E}(XY)=\mathbb{E}\left(F^{-1}(1-U)G^{-1}(U)\right).
	\end{equation} 
\end{lemma}

\begin{theorem}[Sharp Bounds]\label{bounds theorem}
	Let $X_i\sim F_{i}$, $i=1,2$, in which $F_2$ has domain $[\underbar{x}_2,\bar{x}_2]$, $X_2^d\sim G_2,$ $d\in \mathbb{N^+}$, and $U\overset{d}{=}V\sim U[0,1]$ such that $U\perp V$. Let $g^{-1}(y)$ $($resp., $f^{-1}(y)$$)$ be the inverse function of $g(x)=x-F_2(-F_2^{-1}(x))$ $($resp., $f(x)=-g(x)$$)$, and $g'(x)$ $($resp., $f'(x)$$)$ be the derivative of $g(x)$ $($resp., $f(x)$$)$ with respect to $x$. Moreover, denote an indicator function $J=\mathds{1}_{V>\frac{\lvert g'(U_2) \rvert}{\lvert f'(U_2) \rvert+\lvert g'(U_2) \rvert}},$ in which $U_2=F_2(X_2)$.
	\begin{enumerate}[(1)]
		\item Let $\underbar{x}_2\leq 0<\bar{x}_2$ and $\lvert\underbar{x}_2\rvert\leq\bar{x}_2$. $M=\mathbb{E}\left(F_1^{-1}(U)G_2^{-1}(U)\right)$ is obtained when $X_i=F_i^{-1}(U_i)$ with, if $d$ is even,
		\begin{equation}\label{max copula -a<b}
			\begin{aligned}
				U_1&=U, \\
				U_2&=(1-I)\left[Jg^{-1}(U)+(1-J)f^{-1}(U)\right]+IU,
			\end{aligned}
		\end{equation} where $I=\mathds{1}_{U>F_2(-\underbar{x}_2)}$; if $d$ is odd, $U_1=U_2=U$. Moreover, $m=\mathbb{E}\left(F_1^{-1}(1-U)G_2^{-1}(U)\right)$ is obtained when $X_i=F_i^{-1}(U_i)$ with, if $d$ is even,
		\begin{equation}\label{min copula -a<b}
			\begin{aligned}
				U_1&=1-U, \\
				U_2&=(1-I)\left[Jg^{-1}(U)+(1-J)f^{-1}(U)\right]+IU;
			\end{aligned}
		\end{equation} if $d$ is odd, $U_1=1-U$ and $U_2=U$.
		\item Let $\underbar{x}_2< 0\leq\bar{x}_2$ and $\lvert\underbar{x}_2\rvert\geq\bar{x}_2$. $M=\mathbb{E}\left(F_1^{-1}(U)G_2^{-1}(U)\right)$ is obtained when $X_i=F_i^{-1}(U_i)$ with, if $d$ is even,
		\begin{equation}\label{max copula -a>b}
			\begin{aligned}
				U_1&=U, \\
				U_2&=(1-K)\left[Jg^{-1}(U)+(1-J)f^{-1}(U)\right]+K(1-U),
			\end{aligned}
		\end{equation} where $I=\mathds{1}_{U>1-F_2(-\bar{x}_2)}$; if $d$ is odd, $U_1=U_2=U$. Moreover, $m=\mathbb{E}\left(F_1^{-1}(1-U)G_2^{-1}(U)\right)$ is obtained when $X_i=F_i^{-1}(U_i)$ with, if $d$ is even,
		\begin{equation}\label{min copula -a>b}
			\begin{aligned}
				U_1&=1-U, \\
				U_2&=(1-K)\left[Jg^{-1}(U)+(1-J)f^{-1}(U)\right]+K(1-U);
			\end{aligned}
		\end{equation}if $d$ is odd, $U_1=1-U$ and $U_2=U$
	\end{enumerate}
\end{theorem}
\begin{proof}
	We apply Lemma~\ref{max product} and \ref{min product} to prove the theorem. 
	
	(1) For the upper bound and even $d$, the CDF of $X_2^d$ is
	\begin{equation*}
		\begin{aligned}
			G(x)=\mathbb{P}\left(X_2^d\leq x\right)&=\left\{\begin{aligned}
				&\mathbb{P}\left(-x^{\frac{1}{d}}\leq X_2\leq x^{\frac{1}{d}}\right),&\text{ if } x\leq \underbar{x}_2^d,\\
				&\mathbb{P}\left(X_2\leq x^{\frac{1}{d}}\right),&\text{ if } x>\underbar{x}_2^d,
			\end{aligned}
			\right.\\
			&=\left\{\begin{aligned}
				&F_2\left(x^{\frac{1}{d}}\right)-F_2\left(-x^{\frac{1}{d}}\right),&\text{ if } x\leq \underbar{x}_2^d,\\
				&F_2\left(x^{\frac{1}{d}}\right),&\text{ if } x> \underbar{x}_2^d,
			\end{aligned}
			\right.
		\end{aligned}
	\end{equation*} where $x\geq0$. Because of Lemma~\ref{max product}, we have
	$U=F_1(X_1)=U_1,$ and
		\begin{align*}
			U=G(X_2^d)&=\begin{cases}
				F_2\left((X_2^{d})^{\frac{1}{d}}\right)-F_2\left(-(X_2^{d})^{\frac{1}{d}}\right),&\text{ if } X_2^d\leq \underbar{x}_2^d,\\
				F_2\left((X_2^{d})^{\frac{1}{d}}\right),&\text{ if } X_2^d> \underbar{x}_2^d,
			\end{cases}\\
			&=\begin{cases}
				F_2\left(-X_2\right)-F_2\left(X_2\right),&\text{ if }  X_2\leq 0,\\
				F_2\left(X_2\right)-F_2\left(-X_2\right),&\text{ if } 0< X_2\leq \lvert\underbar{x}_2\rvert,\\
				F_2\left(X_2\right),&\text{ if } X_2> \lvert\underbar{x}_2\rvert,
			\end{cases}\\
			&=\begin{cases}
				F_2\left(-F_2^{-1}(U_2)\right)-U_2,&\text{ if } U_2\leq F_2(0),\\
				U_2-F_2\left(-F_2^{-1}(U_2)\right),&\text{ if } F_2(0)< U_2\leq F_2(\lvert\underbar{x}_2\rvert),\\
				U_2,&\text{ if } U_2> F_2(\lvert\underbar{x}_2\rvert).
			\end{cases}
		\end{align*}
	where $U_2=F_2(X_2)$. Let $f(U_2)=F_2\left(-F_2^{-1}(U_2)\right)-U_2$ and $g(U_2)=U_2-F_2\left(-F_2^{-1}(U_2)\right)$. We have
	\begin{equation*}
			U=\left\{\begin{aligned}
				&f(U_2),&\text{ if }& U_2\leq F_2(0),\\
				&g(U_2),&\text{ if }& F_2(0)< U_2\leq F_2(\lvert\underbar{x}_2\rvert),\\
				&U_2,&\text{ if }& U_2> F_2(\lvert\underbar{x}_2\rvert).
			\end{aligned}
			\right.
	\end{equation*}
	Next, we must find the inverse of $U$, i.e., the function of $U_2$. When $U> F_2(\lvert\underbar{x}_2\rvert)$, it is evident that $U_2=U$. When $U\leq F_2(\lvert\underbar{x}_2\rvert)$, $U_2=f^{-1}(U)$ or $U_2=g^{-1}(U)$. Then, we have to find the probabilities of $U_2=f^{-1}(U)$ and $U_2=g^{-1}(U)$. Since the CDF of $U$, $F_U(u)$, is a two-to-one function for $u\leq F_2(\lvert\underbar{x}_2\rvert)$, the change of variables formula gives the following density function for $U$:
	\begin{equation*}
		h_U(u)=\frac{h_U(f^{-1}(u))}{\lvert f'(u_2) \rvert}+\frac{h_U(g^{-1}(u))}{\lvert g'(u_2) \rvert}=\frac{1}{\lvert f'(u_2) \rvert}+\frac{1}{\lvert g'(u_2) \rvert}.
	\end{equation*}Note that $u_2=f^{-1}(u)$ and $u_2=g^{-1}(u)$ are two functions of $u$, but we do not take the derivative with respect to $u$ within $f^{-1}(u)$ and $g^{-1}(u)$. By Bayes' theorem, we have
	\begin{equation*}
		\mathbb{P}(F_U^{-1}(u)=f^{-1}(u)\vert U=u)=\frac{\frac{1}{\lvert f'(u_2) \rvert}}{\frac{1}{\lvert f'(u_2) \rvert}+\frac{1}{\lvert g'(u_2) \rvert}}=\frac{\lvert g'(u_2) \rvert}{\lvert f'(u_2) \rvert+\lvert g'(u_2) \rvert},
	\end{equation*}and
	\begin{equation*}
		\mathbb{P}(F_U^{-1}(u)=g^{-1}(u)\vert U=u)=\frac{\frac{1}{\lvert g'(u_2) \rvert}}{\frac{1}{\lvert f'(u_2) \rvert}+\frac{1}{\lvert g'(u_2) \rvert}}=\frac{\lvert f'(u_2) \rvert}{\lvert f'(u_2) \rvert+\lvert g'(u_2) \rvert}.
	\end{equation*} Hence, the probability of $U_2=f^{-1}(U)$ and $U_2=g^{-1}(U)$ are equal to $$p=\frac{\lvert g'(U_2) \rvert}{\lvert f'(U_2) \rvert+\lvert g'(U_2) \rvert} \text{ and } q=1-p,$$ respectively. Let $J=\mathds{1}_{V>p}$.
	Finally, we obtain
	\begin{equation}\label{U_2}
		U_2=(1-I)\left[Jg^{-1}(U)+(1-J)f^{-1}(U)\right]+IU.
	\end{equation} For the upper bound and odd $d$, $G(x)=\mathbb{P}\left(X_2^d\leq x\right)=F_2(x^{\frac{1}{d}})$, where $x\geq \underbar{x}_2^d$. Thus, $U=G(X_2^d)=F_2(X_2)=U_2$. For the lower bound, Lemma~\ref{min product} derives $U_1=1-U$ and the identical $U_2$ in Equation~\eqref{U_2} or $U_2=U$.
	
	(2) We omit the proof because it is similar to (1). However, it is worth to mention that when $d$ is even, $\underbar{x}_2< 0\leq\bar{x}_2$ and $\lvert\underbar{x}_2\rvert\geq\bar{x}_2$, we have
	\begin{align*}
		U=G(X_2^d)&=\begin{cases}
			1-U_2,&\text{ if } U_2\leq F_2(-\bar{x}_2),\\
			F_2\left(-F_2^{-1}(U_2)\right)-U_2,&\text{ if } F_2(-\bar{x}_2)<U_2\leq F_2(0),\\
			U_2-F_2\left(-F_2^{-1}(U_2)\right),&\text{ if } F_2(0)< U_2,\\
		\end{cases}\\
		&=\begin{cases}
			1-U_2,&\text{ if } U_2\leq F_2(-\bar{x}_2),\\
			f(U_2),&\text{ if } F_2(-\bar{x}_2)<U_2\leq F_2(0),\\
			g(U_2),&\text{ if } F_2(0)< U_2.\\
		\end{cases}
	\end{align*} By inverse the function $U$, we obtain $U_2$ as in Equations~\ref{max copula -a>b} and \ref{min copula -a>b}.
\end{proof}
When $X_2$ is unbounded, i.e., $X_2\in(-\infty, +\infty)$, the representations in \eqref{max copula -a<b}-\eqref{min copula -a>b} still hold. In this case, $I=0$ in \eqref{max copula -a<b}--\eqref{min copula -a>b}. The forms of functions $g^{-1}(U)$ and $f^{-1}(U)$ in Equations~\eqref{max copula -a<b}-\eqref{min copula -a>b} are implicit. However, their expressions, which only depend on the CDF of $X_2$, is easy to derive once $F_2$ is known. Hereafter, propositions for specific marginals and $d=2$ are studied to find explicit copulas.

\begin{proposition}[Non-negative or Non-positive Marginal]\label{non-negative or positive}
	Let $X_i\sim F_{i}$, $i=1,2$, in which $F_2$ has domain $[\underbar{x}_2,\bar{x}_2]$ and $U\sim U[0,1]$.
	\begin{enumerate}[(1)]
		\item Let $\underbar{x}_2\geq 0$. $M$ is attained by a random vector $(X_1,X_2)$, with $X_i=F_i^{-1}(U)$. Moreover, $m$ is attained by a random vector $(X_1,X_2)$, with $X_1=F_1^{-1}(1-U)$ and $X_2=F_2^{-1}(U)$.
		\item Let $\bar{x}_2\leq 0$. $M$ is attained by a random vector $(X_1,X_2)$, with $X_1=F_1^{-1}(1-U)$ and $X_2=F_2^{-1}(U)$. Moreover, $m$ is attained by a random vector $(X_1,X_2)$, with $X_i=F_i^{-1}(U)$.
	\end{enumerate}
\end{proposition}
We omit the proof of Proposition~\ref{non-negative or positive} because its (1) and (2) are direct corollaries of Theorem~\ref{max product}. Copulas to obtain $M$ and $m$ in (1) (resp., (2)) of Proposition~\eqref{non-negative or positive} are special cases of that in Equations~\eqref{max copula -a<b} and \eqref{min copula -a<b} (resp., \eqref{max copula -a>b} and \eqref{min copula -a>b}). When the domain of $F_2$ is non-negative or non-positive, then $I=1$ in \eqref{max copula -a<b}--\eqref{min copula -a>b}. Moreover, $g^{-1}(U)$ and $f^{-1}(U)$ are disappeared for the maximizing and minimizing cases.

\begin{proposition}[Symmetric Marginal with Zero Mean] \label{symmetric marginal}
	Let $X_i\sim F_{i}$, $i=1,2$, in which $F_{2}$ is symmetric with zero mean, $X_2^d\sim G_2$, and $U\overset{d}{=}V\sim U[0,1]$ such that $U\perp V$. The upper bound $M=\mathbb{E}\left(F_1^{-1}(U)G_2^{-1}(U)\right)$ is obtained when $X_i=F_i^{-1}(U_i)$ with, if $d$ is even,
	\begin{equation}\label{max copula in finance}
		\begin{aligned}
			U_1&=U, \\
			U_2&=J\frac{1+U}{2}+(1-J)\frac{1-U}{2},
		\end{aligned}
	\end{equation} where $J=\mathds{1}_{V>\frac{1}{2}}$. Furthermore, the lower bound $m=\mathbb{E}\left(F_1^{-1}(1-U)G_2^{-1}(U)\right)$ is obtained when $X_i=F_i^{-1}(U_i)$ with \begin{equation}\label{min copula in finance}
		\begin{aligned}
			U_1&=1-U, \\
			U_2&=J\frac{1+U}{2}+(1-J)\frac{1-U}{2}.
		\end{aligned}
	\end{equation}
\end{proposition}
\begin{proof} When $F_{2}$ is symmetric with zero mean, $I=0$ in Equations~\eqref{max copula -a<b}--\eqref{min copula -a>b}. For the upper bound and even $d$, we have
	\begin{equation*}
		\begin{aligned}
			g(U_2)&=U_2-F_2(-F_2^{-1}(U_2))=U_2-(1-F_2(F_2^{-1}(U_2)))=2U_2-1,\\
			f(U_2)&=F_2(-F_2^{-1}(U_2))-U_2=1-F_2(F_2^{-1}(U_2))-U_2=1-2U_2,
		\end{aligned}
	\end{equation*}$\Rightarrow g^{-1}(U)=\frac{1+U}{2}$, 
			$f^{-1}(U)=\frac{1-U}{2}$, $g'(U_2)=2$ and $f'(U_2)=-2$. For the lower bound, we have $U_1=1-U$ and the same $U_2$.
\end{proof}

\begin{figure}
	\begin{subfigure}{0.5\textwidth}
		\centering
		\begin{tikzpicture}
			\begin{axis}[
				xlabel={$U_1$},
				ylabel={$U_2$},
				xmin=0, xmax=1.05,
				ymin=0, ymax=1.05,
				axis lines=middle,
				width=\linewidth,
				height=\textwidth,
				xlabel style={at={(current axis.south)}, below=5mm},
				ylabel style={at={(current axis.west)}, xshift=-14mm},
				]
				\addplot[domain=0:1, samples=100, blue] {(1+x)/2};
				\addplot[domain=0:1, samples=100, orange] {(1-x)/2};
				\addplot [dashed] coordinates {(0,0.5) (1,0.5)};
				\addplot [] coordinates {(0,1) (1,1)};
				\addplot [] coordinates {(1,0) (1,1)};
				\node[anchor=south, text=blue] at (axis cs:0.5,0.6) {$g^{-1}(U)$};
				\node[anchor=south, text=orange] at (axis cs:0.5,0.3) {$f^{-1}(U)$};
			\end{axis}
		\end{tikzpicture}
	\end{subfigure}%
	\begin{subfigure}{0.5\textwidth}
		\centering
		\begin{tikzpicture}
			\begin{axis}[
				xlabel={$U_1$},
				xmin=0, xmax=1.05,
				ymin=0, ymax=1.05,
				axis lines=middle,
				width=\linewidth,
				height=\textwidth,
				xlabel style={at={(current axis.south)}, below=5mm},
				]
				\addplot[domain=0:1, samples=100, blue] {1-x/2};
				\addplot[domain=0:1, samples=100, orange] {x/2};
				\addplot [dashed] coordinates {(0,0.5) (1,0.5)};
				\addplot [] coordinates {(0,1) (1,1)};
				\addplot [] coordinates {(1,0) (1,1)};
				\node[anchor=south, text=blue] at (axis cs:0.5,0.6) {$g^{-1}(U)$};
				\node[anchor=south, text=orange] at (axis cs:0.5,0.3) {$f^{-1}(U)$};
			\end{axis}
		\end{tikzpicture}
	\end{subfigure}
	\caption{Support of the copula that maximizes (left panel) resp.\ minimizes (right panel) $\mathbb{E}(X_1X_2^{d})$ where $X_i\sim F_i$, $i=1,2,$ and $d$ is even such that $F_2$ is symmetric with zero mean.}
	\label{opt_copulas_dim2}
\end{figure}
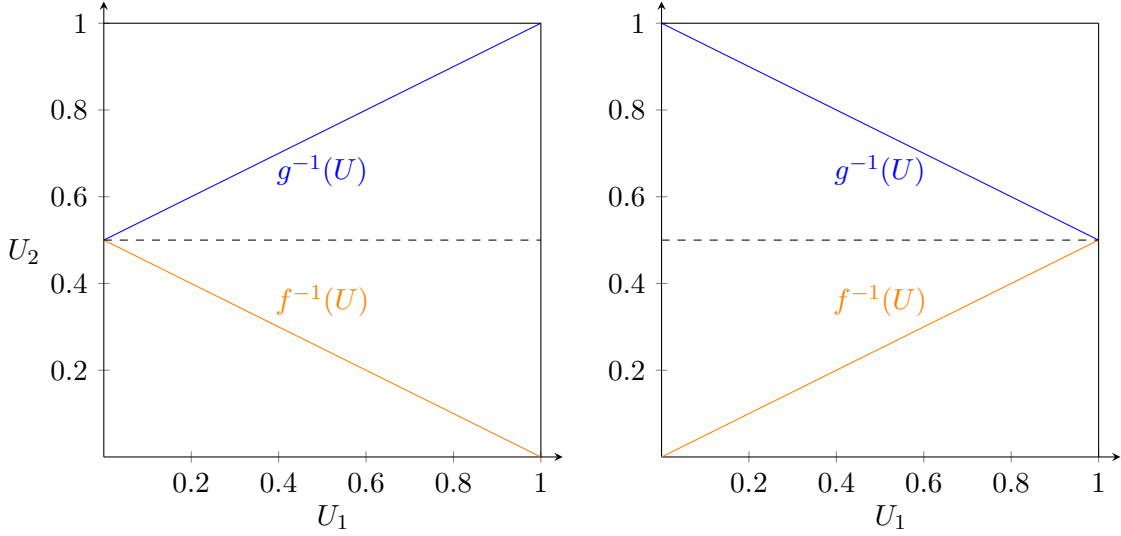
Figure~\ref{opt_copulas_dim2} presents the supports of the copulas in~\eqref{max copula in finance} and \eqref{min copula in finance}. The dashed line is $U_2=F_2(0)=\frac{1}{2}$ for any symmetric distributions $F_2$ with zero means.

The following proposition gives the explicit bounds of $\mathbb{E}[X_1X_2^2]$ and corresponding copulas to attaining them for arbitrary uniform distributions. 

\begin{proposition}[Uniform Marginals]\label{unif marginals}
	Let $X_i\sim U[a_i,b_i]$, $i=1,2$, and $U\overset{d}{=}V\sim U[0,1]$ such that $U\perp V$. 
	\begin{enumerate}[(1)]
		\item Let $a_2\geq0$. The upper bound of $\mathbb{E}[X_1X_2^2]$
		$$M=\frac{1}{12} \left(a_2^2 b_1+2 a_1 a_2 b_2+2 a_2 b_1 b_2+a_1 b_2^2+3 a_1 a_2^2+3 b_1 b_2^2\right)$$ is attained by a random vector $(X_1,X_2)$, with $X_i=a_i+(b_i-a_i)U$. Moreover, the lower bound
		$$m=\frac{1}{12} \left(3 a_2^2 b_1+2 a_1 a_2 b_2+2 a_2 b_1 b_2+3 a_1 b_2^2+a_1 a_2^2+b_1 b_2^2\right)$$ is attained by a random vector $(X_1,X_2)$, with $X_1=a_1+(b_1-a_1)U$ and $X_2=a_2+(b_2-a_2)(1-U)$.
		
		\item Let $b_2\leq0$. The upper bound of $\mathbb{E}[X_1X_2^2]$
		$$M=\frac{1}{12} \left(3 a_2^2 b_1+2 a_1 a_2 b_2+2 a_2 b_1 b_2+3 a_1 b_2^2+a_1 a_2^2+b_1 b_2^2\right)$$ is attained by a random vector $(X_1,X_2)$, with $X_1=a_1+(b_1-a_1)U$ and $X_2=a_2+(b_2-a_2)(1-U)$. Moreover, the lower bound $$m=\frac{1}{12} \left(a_2^2 b_1+2 a_1 a_2 b_2+2 a_2 b_1 b_2+a_1 b_2^2+3 a_1 a_2^2+3 b_1 b_2^2\right)$$ is attained by a random vector $(X_1,X_2)$, with $X_i=a_i+(b_i-a_i)U$.
		
		\item Let $a_2<0<b_2$ and $\lvert a_2\rvert \leq b_2$. The upper bound 
		\begin{equation*}\label{max exp for unif}
			M=a_1\frac{a_2^2+a_2b_2+b_2^2}{3} +(b_1-a_1)\frac{5 a_2^4-4 a_2 b_2^3+3 b_2^4}{12 (b_2-a_2)^2}
		\end{equation*} is obtained when $X_i=a_i+(b_i-a_i)U_i$, in which $U_i$ is in Equation~\eqref{max copula -a<b} with $g^{-1}(U)=\frac{1}{2}U-\frac{a_2}{b_2-a_2}$ and $f^{-1}(U)=-\frac{1}{2}U-\frac{a_2}{b_2-a_2}$. Moreover, the lower bound of $\mathbb{E}[X_1X_2^2]$ \begin{equation*}\label{min exp for unif}
			m=a_1\frac{a_2^2+a_2b_2+b_2^2}{3}+\left(b_1-a_1\right) \frac{-a_2^4-4a_2^3 b_2+b_2^4}{12 \left(b_2-a_2\right)^2}
		\end{equation*} is obtained when $X_i=a_i+(b_i-a_i)U_i$, in which $U_i$ is in Equation~\eqref{min copula -a<b}.
		
		\item Let $a_2<0<b_2$ and $\lvert a_2\rvert > b_2$. The upper bound of $\mathbb{E}[X_1X_2^2]$
		\begin{equation*}
			M=a_1\frac{a_2^2+a_2b_2+b_2^2}{3} +(b_1-a_1)\frac{3a_2^4-4 a_2^3b_2+5b_2^4}{12 (b_2-a_2)^2}
		\end{equation*} is obtained when $X_i=a_i+(b_i-a_i)U_i$, in which $U_i$ is in Equation~\eqref{max copula -a>b} with $g^{-1}(U)=\frac{1}{2}U-\frac{a_2}{b_2-a_2}$ and $f^{-1}(U)=-\frac{1}{2}U-\frac{a_2}{b_2-a_2}$. Moreover, the lower bound \begin{equation*}
			m=a_1\frac{a_2^2+a_2b_2+b_2^2}{3}+\left(b_1-a_1\right) \frac{a_2^4-4 a_2 b_2^3-b_2^4}{12 (b_2-a_2)^2}
		\end{equation*} is obtained when $X_i=a_i+(b_i-a_i)U_i$, in which $U_i$ is in Equation~\eqref{min copula -a>b}.
	\end{enumerate}
\end{proposition}
\begin{proof}
	(1) and (2) can directly be obtained from Proposition~\ref{non-negative or positive}, and $m$ and $M$ are easy to compute. 
	
	(3) When $F_2(x)=\frac{x-a_2}{b_2-a_2}$, $g(U_2)=2U_2+\frac{2a_2}{b_2-a_2}$ and $f(U_2)=-2U_2-\frac{2a_2}{b_2-a_2}$ $\Rightarrow$ $g^{-1}(U)=\frac{1}{2}U-\frac{a_2}{b_2-a_2}$, $f^{-1}(U)=-\frac{1}{2}U-\frac{a_2}{b_2-a_2}$, $g'(U_2)=2$ and $f'(U_2)=-2$ in \eqref{max copula -a<b} and \eqref{min copula -a<b} of Theorem~\ref{risk bounds}. For the upper bound, we have
	\begin{equation*}
		\begin{aligned}
			M=&\mathbb{E}\left(X_1X_2^2\right)\\
			=&a_1\mathbb{E}\left(X_2^2\right) +(b_1-a_1)\mathbb{E}\left((1-I)\frac{(b_2-a_2)^2}{4}U^3+IU((b_2-a_2)U+a_2)^2\right)\\
			=&a_1\frac{a_2^2+a_2b_2+b_2^2}{3} +\\
			&(b_1-a_1)\left[(b_2-a_2)^2\left(\frac{\mathbb{E}\left((1-I)U^3\right)}{4}+\mathbb{E}\left(IU^3\right)\right)+2a_2(b_2-a_2)\mathbb{E}\left(IU^2\right)+a_2^2\mathbb{E}\left(IU\right)\right]\\
			=&a_1\frac{a_2^2+a_2b_2+b_2^2}{3} +(b_1-a_1)\frac{5 a_2^4-4 a_2 b_2^3+3 b_2^4}{12 (b_2-a_2)^2}.
		\end{aligned}
	\end{equation*} Similarly, for the lower bound, we have
	\begin{equation*}
		\begin{aligned}
			m=&\mathbb{E}\left(X_1X_2^2\right)\\
			=&a_1\frac{a_2^2+a_2b_2+b_2^2}{3}+\left(b_1-a_1\right) \frac{-a_2^4-4a_2^3 b_2+b_2^4}{12 \left(b_2-a_2\right)^2}.
		\end{aligned}
	\end{equation*}
	
	(4) We omit the proof because it is similar to (3).
\end{proof}
\begin{figure}
	\begin{subfigure}{0.5\textwidth}
		\centering
		\begin{tikzpicture}
			\begin{axis}[
				xlabel={$U_1$},
				ylabel={$U_2$},
				xmin=0, xmax=1.05,
				ymin=0, ymax=1.05,
				axis lines=middle,
				width=\linewidth,
				height=\textwidth,
				xlabel style={at={(current axis.south)}, below=5mm},
				ylabel style={at={(current axis.west)}, xshift=-14mm},
				]
				\addplot[domain=0:1/2, samples=100, blue] {x/2+1/4};
				\addplot[domain=0:1/2, samples=100, orange] {-x/2+1/4};
				\addplot[domain=1/2:1, samples=100, red] {x};
				\addplot [] coordinates {(0,1) (1,1)};
				\addplot [dashed] coordinates {(0,0.25) (1,0.25)};
				\addplot [dashed] coordinates {(0,0.5) (1,0.5)};
				\addplot [] coordinates {(1,0) (1,1)};
				\addplot [dashed] coordinates {(0.5,0) (0.5,1)};
				\node[anchor=south, text=blue] at (axis cs:0.4,0.3) {$g^{-1}(U)$};
				\node[anchor=south, text=orange] at (axis cs:0.4,0.1) {$f^{-1}(U)$};
				\node[anchor=south, text=red] at (axis cs:0.75,0.8) {$U$};
			\end{axis}
		\end{tikzpicture}
	\end{subfigure}%
	\begin{subfigure}{0.5\textwidth}
		\centering
		\begin{tikzpicture}
			\begin{axis}[
				xlabel={$U_1$},
				xmin=0, xmax=1.05,
				ymin=0, ymax=1.05,
				axis lines=middle,
				width=\linewidth,
				height=\textwidth,
				xlabel style={at={(current axis.south)}, below=5mm},
				]
				\addplot[domain=1/2:1, samples=100, blue] {-x/2+3/4};
				\addplot[domain=1/2:1, samples=100, orange] {x/2-1/4};
				\addplot[domain=0:1/2, samples=100, red] {1-x};
				\addplot [] coordinates {(0,1) (1,1)};
				\addplot [dashed] coordinates {(0,0.25) (1,0.25)};
				\addplot [dashed] coordinates {(0,0.5) (1,0.5)};
				\addplot [] coordinates {(1,0) (1,1)};
				\addplot [dashed] coordinates {(0.5,0) (0.5,1)};
				\node[anchor=south, text=blue] at (axis cs:0.6,0.3) {$g^{-1}(U)$};
				\node[anchor=south, text=orange] at (axis cs:0.6,0.1) {$f^{-1}(U)$};
				\node[anchor=south, text=red] at (axis cs:0.3,0.8) {$1-U$};
			\end{axis}
		\end{tikzpicture}
	\end{subfigure}
	\caption{Support of the copula that maximizes (left panel) resp.\ minimizes (right panel) $\mathbb{E}(X_1X_2^{2})$ where $X_1\sim F_1$ and $X_2\sim U[-1,3]$.}
	\label{opt_copulas_unif}
\end{figure}
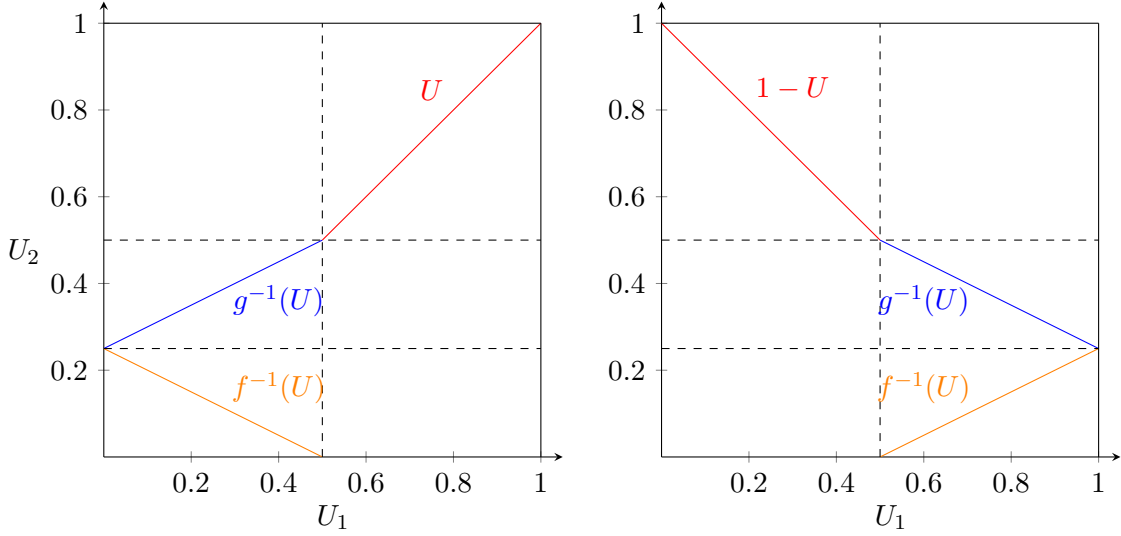
\begin{remark}
	$X_1$ does not affect the maximizing and minimizing copulas to reach $m$ and $M$ in Problems~\eqref{sharp bounds}. The dependence structures to reach them depend on $X_2$ only. However, $X_1$ does affect the magnitudes of $m$ and $M$. We assume $X_1$ be an uniformly distributed random variable in Proposition~\ref{unif marginals} to compute $m$ and $M$ explicitly. Such an assumption is not necessary to prove that dependence structures in Proposition~\ref{unif marginals} maximize and minimize the expectation.
\end{remark}
Figure~\ref{opt_copulas_unif} illustrates the supports of the maximizing and minimizing copulas when $X_1$ be any random variable and $X_2\sim U[-1,3]$. In this case, $I=\mathds{1}_{U>\frac{1}{2}}$, $K=\mathds{1}_{U>\frac{1}{2}}$, $g^{-1}(U)\in\left[\frac{1}{4}, \frac{1}{2}\right]$ and $f^{-1}(U)\in\left[0,\frac{1}{4}\right]$ in (1) of Theorem~\ref{bounds theorem}. 

\begin{figure}
	\begin{subfigure}{0.5\textwidth}
		\centering
		\begin{tikzpicture}
			\begin{axis}[
				xlabel={$U_1$},
				ylabel={$U_2$},
				xmin=0, xmax=1.05,
				ymin=0, ymax=1.05,
				axis lines=middle,
				width=\linewidth,
				height=\textwidth,
				xlabel style={at={(current axis.south)}, below=5mm},
				ylabel style={at={(current axis.west)}, xshift=-14mm},
				]
				\addplot[domain=0.75:1, samples=100, red] {x};
				\addplot[domain=0:0.75, samples=100, blue] {1+(x-sqrt(x^2+1))/2};
				\addplot[domain=0:0.75, samples=100, orange] {1-(x+sqrt(x^2+1))/2};
				\addplot [] coordinates {(0,1) (1,1)};
				\addplot [dashed] coordinates {(0,0.75) (1,0.75)};
				\addplot [dashed] coordinates {(0,0.5) (1,0.5)};
				\addplot [] coordinates {(1,0) (1,1)};
				\addplot [dashed] coordinates {(0.75,0) (0.75,1)};
				\node[anchor=south, text=blue] at (axis cs:0.4,0.53) {$g^{-1}(U)$};
				\node[anchor=south, text=orange] at (axis cs:0.4,0.3) {$f^{-1}(U)$};
				\node[anchor=south, text=red] at (axis cs:0.875,0.875) {$U$};
			\end{axis}
		\end{tikzpicture}
	\end{subfigure}%
	\begin{subfigure}{0.5\textwidth}
		\centering
		\begin{tikzpicture}
			\begin{axis}[
				xlabel={$U_1$},
				xmin=0, xmax=1.05,
				ymin=0, ymax=1.05,
				axis lines=middle,
				width=\linewidth,
				height=\textwidth,
				xlabel style={at={(current axis.south)}, below=5mm},
				]
				\addplot[domain=0:0.25, samples=100, red] {1-x};
				\addplot[domain=0.25:1, samples=100, blue] {1+(1-x-sqrt((1-x)^2+1))/2};
				\addplot[domain=0.25:1, samples=100, orange] {1-(1-x+sqrt((1-x)^2+1))/2};
				\addplot [] coordinates {(0,1) (1,1)};
				\addplot [dashed] coordinates {(0,0.5) (1,0.5)};
				\addplot [dashed] coordinates {(0,0.75) (1,0.75)};
				\addplot [] coordinates {(1,0) (1,1)};
				\addplot [dashed] coordinates {(0.25,0) (0.25,1)};
				\node[anchor=south, text=blue] at (axis cs:0.6,0.53) {$g^{-1}(U)$};
				\node[anchor=south, text=orange] at (axis cs:0.6,0.3) {$f^{-1}(U)$};
				\node[anchor=south, text=red] at (axis cs:0.175,0.875) {$1-U$};
			\end{axis}
		\end{tikzpicture}
	\end{subfigure}
	\caption{Support of the copula that maximizes (left panel) resp.\ minimizes (right panel) $\mathbb{E}(X_1X_2^{d})$ where $X_1\sim F_1$, $X_2+\frac{\ln 2}{\lambda}\sim \text{Exp}(\lambda)$ and $d$ is even.}
	\label{opt_copulas_expon pos}
\end{figure}
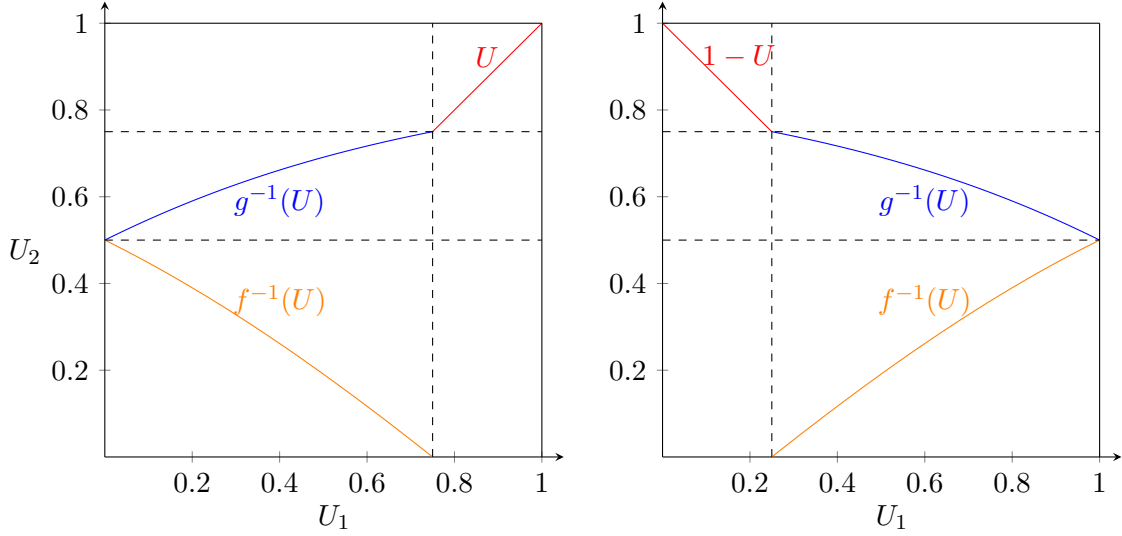

\begin{proposition}[Exponential Distribution]\label{expon dist}
	Let $X_i\sim F_i$, $i=1,2,$ in which $X_2+a\sim \mathrm{Expon}(\lambda)$ with domain $[-a,+\infty)$, $X_2^d\sim G_2$ and $U\overset{d}{=}V\sim U[0,1]$ such that $U\perp V$. Denote an indicator function $J=\mathds{1}_{V>p}$, where \begin{equation}\label{J for expon dist}
		p=\frac{1+\frac{4e^{-2a\lambda}}{\left(U-\sqrt{U^2+4e^{-2a\lambda}}\right)^2}}{2+\frac{4e^{-2a\lambda}}{\left(U+\sqrt{U^2+4e^{-2a\lambda}}\right)^2}+\frac{4e^{-2a\lambda}}{\left(U-\sqrt{U^2+4e^{-2a\lambda}}\right)^2}}.
	\end{equation}
    $M=\mathbb{E}\left(F_1^{-1}(U)G_2^{-1}(U)\right)$ is obtained when $X_i=F_i^{-1}(U_i)$ with, if $d$ is even,
		\begin{equation}\label{expon max copula a>0}
			\begin{aligned}
				U_1&=U, \\
				U_2&=(1-I)\left[J\left(1+\frac{U-\sqrt{U^2+4e^{-2a\lambda}}}{2}\right)+(1-J)\left(1-\frac{U+\sqrt{U^2+4e^{-2a\lambda}}}{2}\right)\right]+IU,
			\end{aligned}
		\end{equation} where $I=\mathds{1}_{U>F_2(a)}$; if $d$ is odd, $U_1=U$ and $U_2=U$. Moreover, $m=\mathbb{E}\left(F_1^{-1}(1-U)G_2^{-1}(U)\right)$ is obtained when $X_i=F_i^{-1}(U_i)$ with, if $d$ is even, $U_1=1-U$ and $U_2$ as in Equations~\eqref{expon max copula a>0}; if $d$ is odd, $U_1=1-U$ and $U_2=U$.
\end{proposition}
We omit the proof of Proposition~\ref{expon dist} since $g^{-1}(U)$ and $f^{-1}(U)$ in (1) of Theorem~\ref{bounds theorem} are easy to compute when $F_2(x)=1-e^{-\lambda (x+a)}$. Figure~\ref{opt_copulas_expon pos} shows the supports of maximizing and minimizing copulas when $X_2+\frac{\ln 2}{\lambda}\sim \mathrm{Expon}(\lambda)$ and $d$ is even. Specifically,
$$g^{-1}(U)=1+\frac{U-\sqrt{U^2+1}}{2}\text{ and }f^{-1}(U)=1-\frac{U+\sqrt{U^2+1}}{2},$$ $I=\mathds{1}_{U>\frac{3}{4}}$, $J=\mathds{1}_{V>p}$ where $p$ in \eqref{J for expon dist} with $a=\frac{\ln 2}{\lambda}$, $g^{-1}(U)\in\left[\frac{1}{2}, \frac{3}{4}\right]$ and $f^{-1}(U)\in\left[0,\frac{1}{2}\right]$. 

\section{Higher moments under dependence uncertainty} \label{first properties of coskewness} 
\cite{jondeau2006optimal} extend the traditional mean-variance framework to incorporate the effect of higher-order centered moments. Specifically, the $(d+1)$-th-order centered mixed moment of $X_1$ and $X_2$ is
\begin{equation*}
	\mu_{d}(X_1,X_2)=\mathbb{E}\left[\left(\frac{X_1-\mu_1}{\sigma_1}\right)\left(\frac{X_2-\mu_2}{\sigma_2}\right)^d\right],
\end{equation*}
where $\mu_i$ and $\sigma_i$, $i=1,2$, are their respective means and standard deviations. $X_1$ can be the return of a risky asset or the loss of an insurance claim, while $X_2$ can be the return of the market or the loss of an overall portfolio. It is clear from the formula of $\mu_{d}(X_1,X_2)$ that the means and standard deviations of the underlying distributions do not affect the comoments. Specifically, we aim at solving the following problems
\begin{subequations}
	\begin{align}
		\underline{S_d}=&\inf \limits_{\ X_i\sim F_i, i=1,2}\mu_{d}(X_1,X_2)\label{mincoskew}, \\
		\overline{S_d}=&\sup \limits_{\ X_i\sim F_i, i=1,2}\mu_{d}(X_1,X_2)\label{maxcoskew}.
	\end{align}
\end{subequations}
Note that solving problems $\underline{S_d}$ and $\overline{S_d}$ under the assumption of $X_i\sim F_i$ is equivalent to solving problems $m$ and $M$ for the case $X_i\sim H_i$ since $X_i\sim F_i \iff Y_i=\frac{X_i-\mu_i}{\sigma_i}\sim H_i$, where the quantile function $H_i^{-1}=\frac{F_i^{-1}-\mu_i}{\sigma_i}$. Thus, we can apply the theory built in the previous section to solve risk bounds on the centered mixed moments $\mu_{d}(X_1,X_2)$ with given marginal distributions $F_1$ and $F_2$, but unknown dependence structure.

\subsection{Risk bounds}
The following proposition and corollary are direct applications of Theorem~\ref{bounds theorem}, Propositions~\ref{symmetric marginal} and \ref{unif marginals}.

\begin{proposition} \label{any marginals}
	Let $X_i\sim F_i$, $i=1,2$, with mean $\mu_i$ and standard deviation $\sigma_i$, and $U\sim U[0,1]$. The maximum centered mixed moments $\overline{S_d}$ of $X_1$ and $X_2$ under dependence uncertainty is given as $\overline{S_d}=\mathbb{E}\left(G^{-1}_{1}(U)G^{-1}_{2}(U)\right)$ where $G_1$ and $G_2$ are the CDFs of $(X_1-\mu_1)/\sigma_1$ and $(X_2-\mu_2)^d/\sigma_2^d$, respectively, and is attained when $X_i=\mu_i+\sigma_iH_i^{-1}(U_i)$ with $U_i$ as in \eqref{max copula -a<b} or \eqref{max copula -a>b}, and $H_i\sim \frac{F_i-\mu_i}{\sigma_i}$; the minimum centered mixed moments
	$\underline{S_d}=\mathbb{E}\left(G^{-1}_{1}(1-U)G^{-1}_{2}(U)\right)$ and is attained when $X_i=\mu_i+\sigma_iH_i^{-1}(U_i)$ with $U_i$ as in \eqref{min copula -a<b} or \eqref{min copula -a>b}. 
\end{proposition}

\begin{corollary}\label{risk bounds for two uniforms}
	Let $X_i\sim F_i$ such that $F_i\sim U[a_i,b_i]$, $i=1,2$. When $d$ is even, $\overline{S_d}=\frac{d3^{\frac{d+1}{2}}}{(d+1)(d+2)}$ and is obtained when $X_i=F_i^{-1}(U_i)$ with $U_i$ as in \eqref{max copula in finance}; $\underline{S_d}=-\frac{d3^{\frac{d+1}{2}}}{(d+1)(d+2)}$ and is obtained when $X_i=F_i^{-1}(U_i)$ with $U_i$ as in \eqref{min copula in finance}. When $d$ is odd, $\overline{S_d}=\frac{3^{\frac{d+1}{2}}}{d+2}$ and is obtained when $X_i=F_i^{-1}(U_i)$ with $U_i=U$; $\underline{S_d}=-\frac{3^{\frac{d+1}{2}}}{d+2}$ and is obtained when $X_i=F_i^{-1}(U_i)$ with $U_1=U$ and $U_2=1-U$.
\end{corollary}

Thanks to Proposition~\ref{any marginals} and Theorem~\ref{bounds theorem}, we can calculate the risk bounds on centered mixed moments for any marginal distributions. We provide some examples for the coskewness hereafter; i.e., $\mu_{2}(X_1,X_2)$. 

\textbf{Normal marginal distributions:} Let  $F_i\sim N(\mu_i,\sigma^2_i)$, $i=1,2$. Standardization of the $F_i$ leads to marginal distributions $H_i \sim N(0,1)$. Hence, an application of Proposition~\ref{symmetric marginal} yields that
$\overline{S_2}=\mathbb{E}\left(\Phi^{-1}(U)G^{-1}(U)\right)=\int_{0}^{1}\Phi^{-1}(u)\Phi^{-1}(\frac{1+u}{2})du$ where $G$ is the CDF of $Z^2$ with $Z\sim N(0, 1)$. Moreover,
$\underline{S_2}=\mathbb{E}\left(\Phi^{-1}(1-U)G^{-1}(U)\right)=-\int_{0}^{1}\Phi^{-1}(u)\Phi^{-1}(\frac{1+u}{2})du.$

\textbf{Exponential marginal distributions:} Let $F_i\sim \mathrm{Expon}(\lambda_i)$, $i=1,2$. After standardization we find that 
$\overline{S_2}=\mathbb{E}\left(H^{-1}(U)G^{-1}(U)\right)$ and $\underline{S_2}=\mathbb{E}\left(H^{-1}(1-U)G^{-1}(U)\right)$ where $G$ is the CDF of $Z^2$ with $Z+1\sim \mathrm{Expon}(1)\sim H$. Expansion yields that
\begin{equation*}
	\begin{aligned}
		\overline{S_2}&=\int_{0}^{b}(-\ln(1-u)-1)\left(\ln\left(\frac{eu+\sqrt{e^2u^2+4}}{2}\right)\right)^2du+\int_{b}^{1}(-\ln(1-u)-1)^3du,\\
		\underline{S_2}&=\int_{0}^{b}(-\ln(u)-1)\left(\ln\left(\frac{eu+\sqrt{e^2u^2+4}}{2}\right)\right)^2du+\int_{b}^{1}(-\ln(u)-1)(\ln(1-u)+1)^2du,
	\end{aligned}
\end{equation*}where $b=1-e^{-2}$.

Similar calculations can also be performed for other marginal distributions. Table \ref{risk bounds table} presents the risk bounds of coskewness $\overline{S_2}$ and $\underline{S_2}$ for some examples of marginal distributions. The risk bounds show $\overline{S_2}=-\underline{S_2}$ for symmetric marginals, while it is not true for asymmetric case.

\begin{table}[!htbp]  
	\centering
	\caption{Risk bounds $\overline{S_2}$ and $\underline{S_2}$. The distributions of $X_i$, $i=1,2,$ denoted as $F_i$, are specified in the first column. In the case of Student't, $\sqrt{\frac{\nu_i-2}{\nu_i}}X_i\sim H_i$, the degrees of freedom $\nu_i>3$, and $a=\frac{\nu_2-2}{\nu_2}\sqrt{\frac{\nu_1-2}{\nu_1}}$ such that $\nu_i>2$. For Laplace distributions, $F\sim \mathrm{Laplace}(0, 1)$. In the case of exponential, $b=1-e^{-2}$, $f(x)=\left(\ln\left(\frac{ex+\sqrt{e^2x^2+4}}{2}\right)\right)^2$, $g(x)=-\ln(1-x)-1$ and $h(x)=-\ln(x)-1$.
	}
	\begin{tabular}{|c|c|c|}
		\hline
		Marginal Distributions&Minimum Coskewness&Maximum Coskewness\\
		\hline
		$\mathrm{N}(\mu_i, \sigma_i^2)$& $-\int_{0}^{1}\Phi^{-1}(u)\Phi^{-1}(\frac{1+u}{2})du$ & $\int_{0}^{1}\Phi^{-1}(u)\Phi^{-1}(\frac{1+u}{2})du$\\
		\hline
		$\mathrm{Student}(\nu_i)$ & $-a\int_{0}^{1}H_1^{-1}(u)H_2^{-1}(\frac{1+u}{2})du$ & $a\int_{0}^{1}H_1^{-1}(u)H_2^{-1}(\frac{1+u}{2})du$ \\
		\hline
		$\mathrm{Laplace}(\mu_i, b_i)$ & $-\int_{0}^{1}F^{-1}(u)F^{-1}(\frac{1+u}{2})du$ & $\int_{0}^{1}F^{-1}(u)F^{-1}(\frac{1+u}{2})du$ \\
		\hline
		$\mathrm{Expon}(\lambda_i)$ & $\int_{0}^{b}h(u)f(u)du+\int_{b}^{1}h(u)g(u)^2du$ & $\int_{0}^{b}g(u)f(u)du+\int_{b}^{1}g(u)^3du$ \\
		\hline
	\end{tabular}
	\label{risk bounds table}
\end{table}

\subsection{Standardized rank coefficients}
We define a series of new risk measures which will be useful in understanding the centered mixed moments. In the same spirit of Definition 4.1 in \cite{bernard2023coskewness} and Spearman correlation, we propose standardized rank coefficients hereafter.
\begin{definition}[Standardized Rank Coefficients]\label{standardized rank coskewness in finance}
	Let $X_i\sim F_i$, $i=1,2$. The standardized rank coefficients of $X_1$ and $X_2$, denoted as $RS_d(X_1,X_2)$, is defined by 
	\begin{equation*} 
		RS_d(X_1,X_2)=\frac{2^{d+1}(d+1)(d+2)}{d}\mathbb{E}\left[\left(F_{1}(X_1)-\frac{1}{2}\right)\left(F_{2}(X_2)-\frac{1}{2}\right)^d\right],
	\end{equation*}if $d$ is even; otherwise,
	\begin{equation*} 
		RS_d(X_1,X_2)=2^{d+1}(d+2)\mathbb{E}\left[\left(F_{1}(X_1)-\frac{1}{2}\right)\left(F_{2}(X_2)-\frac{1}{2}\right)^d\right].
	\end{equation*}
\end{definition}
\begin{proposition}
	The standard rank coefficients $RS_d(X_1,X_2)$ satisfies the following properties:
	\begin{enumerate}[(1)]
		\item $-1\leq RS_d(X_1,X_2)\leq 1$.
		\item The upper bound of 1 is obtained when $X_i=F_i^{-1}(U_i)$ with $U_i$ as in \eqref{max copula in finance} if $d$ is even; otherwise, $U_i=U$. The lower bound of -1 is obtained when $X_i=F_i^{-1}(U_i)$ with $U_i$ as in \eqref{min copula in finance} if $d$ is even; otherwise, $U_1=1-U$ and $U_2=U$.
		\item It is invariant under strictly increasing transformation, i.e., for arbitrary strictly increasing functions $f_i$, $i=1,2$, $RS_d(X_1,X_2)= RS_d(f_1(X_1),f_2(X_2))$.
	\end{enumerate}
\end{proposition}
These properties follow from Propositions~\ref{symmetric marginal}, \ref{unif marginals} and Corollary~\ref{risk bounds for two uniforms}. \cite{bucher2017testing} define the non-standardized rank coskewness to propose a new measure of asymmetry for dependence. However, our proposed standardized rank coefficients are easier to interpret and more elegant than the non-standardized one since we fix the range between -1 and 1. 

Next section, we study the effect of these higher-order moments in three critical areas of actuarial science.

\section{The effect of higher-order moments}\label{po ap}
In this section, we examine how higher-order dependence measures, such as coskewness, affect insurance capital requirements, risk capital allocation rules, and the pricing of life annuities.
 
\subsection{Preliminaries}
We consider two random variables, $X_1\sim \mathrm{Expon}(\lambda_1)$ and $X_2\sim \mathrm{Expon}(\lambda_2)$, as the exponential distribution is widely used as a benchmark in actuarial science, particularly in modeling mortality in life insurance and claim amounts in auto insurance.

The dependence structure between $X_1$ and $X_2$ is modeled using the mixture copula $C_{\lambda}$, as defined in the following, based on Propositions~\ref{non-negative or positive}, \ref{expon dist} and Theorem~\ref{sharp bounds}; see a similar notion in \cite{bernard2024modeling}.
\begin{definition}[Copula $C_{\lambda}$]\label{mix copula defi dim 2}
	Let $X_i\sim \mathrm{Expon}(\lambda_i)$, $i=1,2$, $U\overset{d}{=}V\sim U[0,1]$ such that $U\perp V$, $B\sim Bernoulli(\lambda)$ such that $B$ is independent of $X_i$, $U$ and $V$, and $\lambda\in [0,1]$. The dependence structure of $X_1=-\frac{\ln(1-U_{1,\lambda})}{\lambda_1}$, and $X_2=-\frac{\ln(1-U_2)}{\lambda_2}$ is called a mixture copula $C_{\lambda}$ when the bivariate random vector $(U_{1,\lambda}, U_2)$ is given as, if $d$ is even,
	\begin{equation}\label{Exp copula for cosk}
		\begin{aligned}
			U_{1,\lambda}&=BU+(1-B)(1-U),\\
			U_2&=(1-I)\left[J\left(1+\frac{U-\sqrt{U^2+4e^{-2}}}{2}\right)+(1-J)\left(1-\frac{U+\sqrt{U^2+4e^{-2}}}{2}\right)\right]+IU,
		\end{aligned}
	\end{equation} where $I=\mathds{1}_{U>1-e^{-2}}$ and $J=\mathds{1}_{V>p}$, with \begin{equation*}
		p=\frac{\frac{e^{2}}{4}+\frac{1}{\left(U-\sqrt{U^2+4e^{-2}}\right)^2}}{\frac{e^{2}}{2}+\frac{1}{\left(U+\sqrt{U^2+4e^{-2}}\right)^2}+\frac{1}{\left(U-\sqrt{U^2+4e^{-2}}\right)^2}};
	\end{equation*} otherwise, $U_{1,\lambda}=BU+(1-B)(1-U)$ and $U_2=U$.
\end{definition}

The standardized centered mixed moments of $X_1$ and $X_2$, $\mu_{d}(X_1,X_2)$, are given by
\begin{equation*}
	\begin{aligned}
		\mu_{2k-1}(X_1,X_2)
		=&\mathbb{E}\left[(-\ln(1-U_{1,\lambda})-1)\left(-\ln(1-U)-1\right)^{2k-1}\right] \\
		=&\mathbb{E}\left[(-\ln(U)-1)\left(-\ln(1-U)-1\right)^{2k-1}\right](1-\lambda)+\mathbb{E}\left[\left(-\ln(1-U)-1\right)^{2k}\right]\lambda\\
		=&\underline{S_{2k-1}}(1-\lambda)+\overline{S_{2k-1}}\lambda =\underline{S_{2k-1}}+\lambda \left(\overline{S_{2k-1}}-\underline{S_{2k-1}}\right),
	\end{aligned}
\end{equation*}and
\begin{equation*}
	\begin{aligned}
		\mu_{2k}(X_1,X_2)=&\mathbb{E}\left[(-\ln(1-U_{1,\lambda})-1)\left(-\ln(1-U_2)-1\right)^{2k}\right] \\
		=&\mathbb{E}\left[(-\ln(U)-1)\left(-\ln(1-U_2)-1\right)^{2k}\right](1-\lambda)\\
		&+\mathbb{E}\left[(-\ln(1-U)-1)\left(-\ln(1-U_2)-1\right)^{2k}\right]\lambda\\
		=&\underline{S_{2k}}(1-\lambda)+\overline{S_{2k}}\lambda=\underline{S_{2k}}+\lambda (\overline{S_{2k}}-\underline{S_{2k}}).
	\end{aligned}
\end{equation*}It is evident that $\mu_{d}(X_1,X_2)$ is a linear function of $\lambda \in [0, 1]$, and remains invariant under changes to marginal means and standard deviations. By varying the parameter $\lambda$, one can evaluate $\mu_{d}(X_1,X_2)$ from its minimum ($\lambda = 0$) to maximum ($\lambda = 1$). For a detailed study of dependence uncertainty in standardized centered mixed moments involving more than two variables, see \cite{bernard2023coskewness}.

Table~\ref{cosk as a function of l} presents numerical values of $\mu_{d}(X_1,X_2)$ and $RS_{d}(X_1,X_2)$ for $d=1,2,3,4,$ where $X_1 \sim \text{Expon}(\lambda_1)$ and $X_2 \sim \text{Expon}(\lambda_2)$. These values are computed via Monte Carlo simulation by generating $n = 10^8$ independent samples $u_i$, $i = 1, 2, \ldots, n$, from the standard uniform distribution and setting $U = (u_1, u_2, \ldots, u_n)$. Note that the mixed moments $\mu_{d}(X_1,X_2)$, for $d = 1, 2, 3, 4$, correspond respectively to Pearson correlation, coskewness, cokurtosis, and cohyperskewness, while $RS_{d}(X_1,X_2)$ are their corresponding standardized rank coefficients. Specifically, $\mu_{1}(X_1,X_2) = -0.65$ illustrates a known limitation of the Pearson correlation: it is impacted by the marginal distributions, and hence its theoretical lower bound of -1 is not always sharp; see, e.g., \cite{embrechts2002correlation}. Moreover, values of $RS_{d}(X_1,X_2)$ in Table~\ref{cosk as a function of l} illustrate that they are are easier to interpret and more elegant than $\mu_{d}(X_1,X_2)$.

\begin{table}[!h]
	\centering 
	\caption{$\mu_{d}(X_1,X_2)$ and $RS_{d}(X_1,X_2)$, where $d=1,2,3,4,$ $X_1\sim \mathrm{Expon}(\lambda_1)$ and $X_2\sim \mathrm{Expon}(\lambda_2)$.}
	\begin{tabularx}{\textwidth}{*{6}{X}}
		\toprule
		$\lambda$& 0 & 0.25 & 0.5 & 0.75 & 1.00\\
		\midrule
		$\mu_{1}(X_1,X_2)$& -0.65 & -0.23 & 0.18 & 0.59  & 1.00\\
		$\mu_{2}(X_1,X_2)$& -0.81 & -0.06 & 0.69 & 1.45  & 2.21\\
		$\mu_{3}(X_1,X_2)$& -2.43 & 0.43 & 3.28  & 6.14  & 9.00\\
		$\mu_{4}(X_1,X_2)$& -8.83 & 4.39 & 17.71 & 30.87 & 43.99\\
		\midrule
		$RS_{1}(X_1,X_2)$& -1 & -0.50 & 0 & 0.50 & 1\\
		$RS_{2}(X_1,X_2)$& -0.92 & -0.46 & 0 & 0.46 & 0.92\\
		$RS_{3}(X_1,X_2)$& -1 & -0.50 & 0 & 0.50 & 1\\
		$RS_{4}(X_1,X_2)$& -0.94 & -0.47 & 0 & 0.47 & 0.94\\
		\bottomrule
	\end{tabularx} 
	\label{cosk as a function of l}
\end{table}

In Sections~\ref{insur capital req} and \ref{capital risk ar}, we interpret $X_1 := L_1$ and $X_2 := L_2$ as random losses from a new business line and an existing loss portfolio composed of multiple lines of business held by an insurance company. In Section~\ref{pricing life annuit}, we model the future lifetimes of a married couple as $X_1 := T_x$ and $X_2 := T_y$.

In the subsections that follow, we utilize different values of $\lambda$ to demonstrate the magnitude of standardized centered mixed moments, with particular emphasis on $\mu_{2k}(X_1,X_2)$ as measures of dependence. We focus on these higher-order moments to bridge the gap in the literature, while omitting detailed discussion of $\mu_{2k-1}(X_1,X_2)$, such as Pearson correlation, which is well-established.

\subsection{Insurance capital requirement}\label{insur capital req}
In this subsection, we investigate the impact of $(2k-1)$-th-order standardized centered mixed moments, denoted $\mu_{2k}(X_1,X_2)$, on the capital requirement for potential losses faced by insurance institutions subject to global regulatory frameworks. The capital requirement is measured using ES - also referred to as conditional value-at-risk (CVaR), tail value-at-risk (TVaR), or average value-at-risk (AVaR) - due to its critical role in financial regulations such as Solvency III and Basel III.

The ES risk measure for aggregate portfolio risk, at level $p \in (0, 1)$, is defined as
\begin{equation*}
	\label{ES-intro-def}
	\mathrm{ES}_p(S)=\mathbb{E}[S\vert S> \rm{VaR}_p(S)],
\end{equation*} where $\text{VaR}_p(X)$ is the value-at-risk (VaR) of $S=L_1+L_2$ at the prudence level $p$, given by
$\text{VaR}_p(X)=\inf\{x\in\mathbb{R}: F_X(x)\geq p\}$
\citep[see, e.g.,][]{denuit2006actuarial}. 

The losses from a new business line and an existing portfolio are modeled as $L_1=-\frac{\ln(1-U_{1,\lambda})}{\lambda_1}$, and $L_2=-\frac{\ln(1-U_2)}{\lambda_2},$ where $U_{1,\lambda}$ and $U_2$ are defined as in Equations~\eqref{Exp copula for cosk}. Under this specification, the ES of the aggregate loss $S = L_1 + L_2$ is given by
	\begin{equation}\label{ES formula}
		\mathrm{ES}_p(S)=\frac{1}{(1-p)\lambda_1\lambda_2}\int_{p}^{1}\rm{VaR}_q(-\ln((1-U_{1,\lambda})^{\lambda_2}(1-U_2)^{\lambda_1})dq,
	\end{equation} where $U_{1,\lambda}$ and $U_2$ are as defined in \eqref{Exp copula for cosk}.

An explicit analytical expression for $\mathrm{ES}_p(S)$ is generally not available. Hence, we compute it numerically using Equation~\eqref{ES formula}. Specifically, we simulate $n = 10^6$ independent samples $u_i$, $i = 1, 2, \ldots, n$, from the standard uniform distribution to obtain approximations of $\mathrm{ES}_p(S)$.

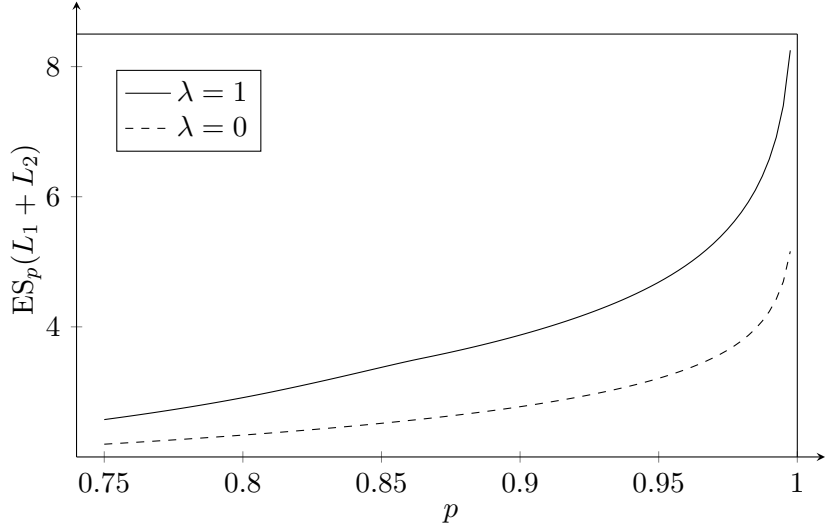
\begin{figure}
	\centering
	\begin{tikzpicture}
	\begin{axis}[
		xlabel={$p$},
		ylabel={$\mathrm{ES}_p(L_1+L_2)$},
		xmin=0.74, xmax=1.01,
		ymin=2, ymax=9,
		axis lines=middle,
		width=0.75\textwidth,
		height=0.5\textwidth,
		xlabel style={at={(current axis.south)}, below=5mm},
		ylabel style={at={(current axis.west)}, rotate=90, yshift=10mm},
		xtick={0.75, 0.8, 0.85, 0.9, 0.95, 1},
		legend style={at={(0.15,0.85)}, anchor=north}
		]
		\addplot[] coordinates {
			(0.75, 2.574392765651651)
			(0.7525252525252525, 2.5887752314813284)
			(0.7550505050505051, 2.6034498908412202)
			(0.7575757575757576, 2.618309351471617)
			(0.76010101010101, 2.633472574723997)
			(0.7626262626262627, 2.6488277241957983)
			(0.7651515151515151, 2.6645017026760196)
			(0.7676767676767677, 2.6803844742299443)
			(0.7702020202020202, 2.6966035431521944)
			(0.7727272727272727, 2.713040628926356)
			(0.7752525252525253, 2.729839592478599)
			(0.7777777777777778, 2.7469480494538114)
			(0.7803030303030303, 2.7642993913671154)
			(0.7828282828282829, 2.7820431950499565)
			(0.7853535353535354, 2.8000450764342832)
			(0.7878787878787878, 2.8184553086074904)
			(0.7904040404040404, 2.8371374083329903)
			(0.7929292929292929, 2.856257526679098)
			(0.7954545454545454, 2.875677857106602)
			(0.797979797979798, 2.8955613748816873)
			(0.8005050505050505, 2.9157619510077772)
			(0.803030303030303, 2.9364590878691867)
			(0.8055555555555556, 2.9575039542636077)
			(0.8080808080808081, 2.9790814854447825)
			(0.8106060606060606, 3.001036900462024)
			(0.8131313131313131, 3.0234711955044595)
			(0.8156565656565656, 3.0460950287069437)
			(0.8181818181818182, 3.0691018350943597)
			(0.8207070707070707, 3.0922933755335245)
			(0.8232323232323232, 3.1158379028460477)
			(0.8257575757575758, 3.1395332560483786)
			(0.8282828282828283, 3.1636074036554462)
			(0.8308080808080808, 3.1878200155738803)
			(0.8333333333333334, 3.212335052428775)
			(0.8358585858585859, 3.237096873415575)
			(0.8383838383838385, 3.2619193715768096)
			(0.8409090909090909, 3.2869428315162543)
			(0.8434343434343434, 3.312003634408752)
			(0.845959595959596, 3.33718375452194)
			(0.8484848484848485, 3.362252597632121)
			(0.851010101010101, 3.3873091677514817)
			(0.8535353535353536, 3.4120993474869246)
			(0.8560606060606061, 3.43685632316574)
			(0.8585858585858586, 3.461179616788156)
			(0.8611111111111112, 3.4848965703755326)
			(0.8636363636363636, 3.5076653271236187)
			(0.8661616161616161, 3.529831710181565)
			(0.8686868686868687, 3.5522222319585364)
			(0.8712121212121212, 3.5751316487622913)
			(0.8737373737373737, 3.598380738108722)
			(0.8762626262626263, 3.622220328167238)
			(0.8787878787878788, 3.64648117572736)
			(0.8813131313131313, 3.671391315926982)
			(0.8838383838383839, 3.6967341082151552)
			(0.8863636363636364, 3.722723169715982)
			(0.8888888888888888, 3.7493193035816863)
			(0.8914141414141414, 3.7763883443031117)
			(0.8939393939393939, 3.8041864852749185)
			(0.8964646464646464, 3.832570625600195)
			(0.898989898989899, 3.861737873909835)
			(0.9015151515151515, 3.891508615846554)
			(0.9040404040404041, 3.9221689070702843)
			(0.9065656565656566, 3.9535830555069755)
			(0.9090909090909092, 3.9859504094948153)
			(0.9116161616161617, 4.0190369692812125)
			(0.9141414141414141, 4.053186626861583)
			(0.9166666666666667, 4.088212503448993)
			(0.9191919191919192, 4.124450645297285)
			(0.9217171717171717, 4.16159085161606)
			(0.9242424242424243, 4.200091648027745)
			(0.9267676767676768, 4.23975237351775)
			(0.9292929292929293, 4.280851144271067)
			(0.9318181818181819, 4.323238245005597)
			(0.9343434343434344, 4.367389818381912)
			(0.9368686868686869, 4.413115560023723)
			(0.9393939393939394, 4.461057672298285)
			(0.9419191919191919, 4.510909286979564)
			(0.9444444444444444, 4.563050833023678)
			(0.946969696969697, 4.61757241113435)
			(0.9494949494949495, 4.674388572000582)
			(0.952020202020202, 4.734591391217915)
			(0.9545454545454546, 4.7981057041066935)
			(0.9570707070707071, 4.86553265024919)
			(0.9595959595959596, 4.936915130773396)
			(0.9621212121212122, 5.013261391800454)
			(0.9646464646464646, 5.094716117553982)
			(0.9671717171717171, 5.18309252050652)
			(0.9696969696969697, 5.278275679176832)
			(0.9722222222222222, 5.38230690307645)
			(0.9747474747474748, 5.495729741142369)
			(0.9772727272727273, 5.6213199280060735)
			(0.9797979797979798, 5.760336148805138)
			(0.9823232323232324, 5.918171080207672)
			(0.9848484848484849, 6.09952798357845)
			(0.9873737373737375, 6.3155764514715464)
			(0.98989898989899, 6.577569512418134)
			(0.9924242424242424, 6.91534519887599)
			(0.994949494949495, 7.398931652512849)
			(0.9974747474747475, 8.250925370266327)
		};
		\addlegendentry{$\lambda=1$}
		\addplot[dashed] coordinates {(0.75, 2.19621200648142)
			(0.7525252525252525, 2.2025548316119643)
			(0.7550505050505051, 2.2089935194266377)
			(0.7575757575757576, 2.2154721922762777)
			(0.76010101010101, 2.222046509110961)
			(0.7626262626262627, 2.2286665521144964)
			(0.7651515151515151, 2.2353820653476846)
			(0.7676767676767677, 2.2421498076231483)
			(0.7702020202020202, 2.249023181177517)
			(0.7727272727272727, 2.255943007203627)
			(0.7752525252525253, 2.2629619079859418)
			(0.7777777777777778, 2.2700605577817776)
			(0.7803030303030303, 2.277214268419676)
			(0.7828282828282829, 2.2844799202836197)
			(0.7853535353535354, 2.2917979188450763)
			(0.7878787878787878, 2.299228150447943)
			(0.7904040404040404, 2.3067158382482704)
			(0.7929292929292929, 2.31433488497918)
			(0.7954545454545454, 2.32202425972498)
			(0.797979797979798, 2.32984213830139)
			(0.8005050505050505, 2.3377249858081064)
			(0.803030303030303, 2.3457377874526717)
			(0.8055555555555556, 2.35381897700668)
			(0.8080808080808081, 2.3620425674689245)
			(0.8106060606060606, 2.3703542703970024)
			(0.8131313131313131, 2.3788233986907152)
			(0.8156565656565656, 2.387374017898563)
			(0.8181818181818182, 2.396071482636278)
			(0.8207070707070707, 2.40485866217102)
			(0.8232323232323232, 2.4138150318870815)
			(0.8257575757575758, 2.4228660802290283)
			(0.8282828282828283, 2.4320907950498407)
			(0.8308080808080808, 2.441409559019293)
			(0.8333333333333334, 2.450908189160835)
			(0.8358585858585859, 2.4605455416960593)
			(0.8383838383838385, 2.470299276570524)
			(0.8409090909090909, 2.480249243453267)
			(0.8434343434343434, 2.4903045667631085)
			(0.845959595959596, 2.50056410361967)
			(0.8484848484848485, 2.5109599819784694)
			(0.851010101010101, 2.5215778940423026)
			(0.8535353535353536, 2.532335031890507)
			(0.8560606060606061, 2.5433093288422524)
			(0.8585858585858586, 2.5544251646237)
			(0.8611111111111112, 2.5657787517147375)
			(0.8636363636363636, 2.5772908181923295)
			(0.8661616161616161, 2.5890526416236757)
			(0.8686868686868687, 2.600995447708089)
			(0.8712121212121212, 2.6132377255249746)
			(0.8737373737373737, 2.6256674031050538)
			(0.8762626262626263, 2.638396396084835)
			(0.8787878787878788, 2.6513215099359497)
			(0.8813131313131313, 2.664581378088544)
			(0.8838383838383839, 2.6780878048808012)
			(0.8863636363636364, 2.6919669986601527)
			(0.8888888888888888, 2.706154864795416)
			(0.8914141414141414, 2.720590565081934)
			(0.8939393939393939, 2.7354173102384585)
			(0.8964646464646464, 2.7505387520795517)
			(0.898989898989899, 2.76608750559924)
			(0.9015151515151515, 2.781985787652118)
			(0.9040404040404041, 2.79838037882309)
			(0.9065656565656566, 2.8151687580550253)
			(0.9090909090909092, 2.832487516550328)
			(0.9116161616161617, 2.850197251818553)
			(0.9141414141414141, 2.8685061333554818)
			(0.9166666666666667, 2.8872864976030415)
			(0.9191919191919192, 2.90674210787545)
			(0.9217171717171717, 2.9267139815636867)
			(0.9242424242424243, 2.947409656844476)
			(0.9267676767676768, 2.9686424837772107)
			(0.9292929292929293, 2.9906586759763965)
			(0.9318181818181819, 3.0133885229464443)
			(0.9343434343434344, 3.03706071256881)
			(0.9368686868686869, 3.0615655840346405)
			(0.9393939393939394, 3.0872238419993447)
			(0.9419191919191919, 3.1138961109024983)
			(0.9444444444444444, 3.1419633701346705)
			(0.946969696969697, 3.1713972561588473)
			(0.9494949494949495, 3.202107709458314)
			(0.952020202020202, 3.234467771294819)
			(0.9545454545454546, 3.2685635736714347)
			(0.9570707070707071, 3.304721237942466)
			(0.9595959595959596, 3.342829128784875)
			(0.9621212121212122, 3.383632717217896)
			(0.9646464646464646, 3.4271927262152584)
			(0.9671717171717171, 3.4740437408663034)
			(0.9696969696969697, 3.52453709521331)
			(0.9722222222222222, 3.579519202527818)
			(0.9747474747474748, 3.639505372443441)
			(0.9772727272727273, 3.706767392949864)
			(0.9797979797979798, 3.782264636069756)
			(0.9823232323232324, 3.8685021427058204)
			(0.9848484848484849, 3.9688883057377358)
			(0.9873737373737375, 4.088739326503105)
			(0.98989898989899, 4.235655776497365)
			(0.9924242424242424, 4.427782823294799)
			(0.994949494949495, 4.693331460621278)
			(0.9974747474747475, 5.160516516992977)
		};
		\addlegendentry{$\lambda=0$}
		\addplot [] coordinates {(0,8.5) (1,8.5)};
		\addplot [] coordinates {(1,0) (1,8.5)};
	\end{axis}
\end{tikzpicture}
	\caption{$\mathrm{ES}_p(L_1+L_2)$ with $\overline{S_{2k}}$ and $\underline{S_{2k}}$, where $L_1\sim \mathrm{Expon}(1.5)$ and $L_2\sim \mathrm{Expon}(2)$, as functions of the prudence level $p\in[0.75, 1)$.}
	\label{es with max and min m}
\end{figure}

In Figure~\ref{es with max and min m}, we plot $\mathrm{ES}_p(L_1 + L_2)$ under the scenarios where $\mu_{2k}(X_1,X_2)$ attains its minimum ($\lambda = 0$) and maximum ($\lambda = 1$) values, as functions of the prudence level $p \in [0.75, 1)$. It is evident that the ES corresponding to the maximum value of $\mu_{2k}(X_1,X_2)$ is consistently higher than that corresponding to the minimum. Furthermore, this difference becomes more pronounced as the prudence level increases.

To further explore this relationship, we present in Panels A–E of Table~\ref{es with different m} the capital requirement values for various combinations of $\lambda_1$, $\lambda_2$, and $p$. Although the growth rates differ across parameter settings, ES increases monotonically with $\mu_{2k}(X_1,X_2)$. Notably, when $\lambda_1$ and $\lambda_2$ increase, the rate of change in ES decreases, whereas higher prudence levels amplify the sensitivity of ES to changes in $\mu_{2k}(X_1,X_2)$. These observations align with the trends illustrated in Figure~\ref{es with max and min m}.
	
\begin{table}[!h]
	\centering 
	\caption{The effect of $\mu_{2k}(X_1,X_2)$ on $\mathrm{ES}_p(L_1+L_2)$, where $L_1\sim \mathrm{Expon}(\lambda_1)$ and $L_2\sim \mathrm{Expon}(\lambda_2)$.}
	\begin{tabularx}{\textwidth}{*{6}{X}}
		\toprule
		$\lambda$& 0 & 0.25 & 0.5 & 0.75 & 1.00\\
		\midrule
		\multicolumn{6}{l}{Panel A: benchmark case ($\lambda_1=\lambda_2=1$ and $p=0.95$)} \\
		$\mathrm{ES}_p(L_1+L_2)$ & 5.32 & 6.19 & 6.83 & 7.5 & 7.96 \\
		\midrule
		\multicolumn{6}{l}{Panel B: changing rate parameter for $L_1$ ($\lambda_1=2,$ $\lambda_2=1$ and $p=0.95$)}\\
		$\mathrm{ES}_p(L_1+L_2)$ & 2.07 & 2.36 & 2.62 & 2.8 & 3.0 \\
		\midrule
		\multicolumn{6}{l}{Panel C: changing rate parameter for $L_2$ ($\lambda_1=1,$ $\lambda_2=2$ and $p=0.95$)}\\
		$\mathrm{ES}_p(L_1+L_2)$ & 2.24 & 2.49 & 2.67 & 2.86 & 3.0 \\
		\midrule
		\multicolumn{6}{l}{Panel D: reducing the prudence level ($\lambda_1=\lambda_2=1$ and $p=0.90$)}\\
		$\mathrm{ES}_p(L_1+L_2)$ & 4.65 & 5.16 & 5.73 & 6.17 & 6.62 \\
		\midrule
		\multicolumn{6}{l}{Panel E: rising the prudence level ($\lambda_1=\lambda_2=1$ and $p=0.99$)}\\
		$\mathrm{ES}_p(L_1+L_2)$ & 6.96 & 8.77 & 9.86 & 10.76 & 11.26 \\
		\bottomrule
	\end{tabularx} 
	\label{es with different m}
\end{table}
The findings in this subsection offer practical insights for actuaries considering the integration of a new business line into an existing portfolio. For example, to reduce capital requirements in compliance with regulatory constraints, an insurer may opt to introduce a new product line that exhibits relatively low coskewness with the existing portfolio, thereby reducing the overall dependence risk as captured by $S(L_1, L_2)$.  

\subsection{Risk capital allocation}\label{capital risk ar}
In this subsection, we examine how the $(2k-1)$-th-order standardized centered mixed moments $\mu_{2k}(X_1,X_2)$ affect the contribution of an individual risk to the overall systemic risk in the market. The MES risk measure is selected to quantify this contribution; see, for example, \cite{acharya2017measuring} for an introduction to MES. Like ES, MES is also a critical component in Basel III and Solvency III.

The MES of an individual risk $L_1 \sim F_1$ with respect to the aggregate (systemic) risk $S=L_1+L_2 \sim F$, at a given prudence level $p$, is defined as
\begin{equation*}
	\label{MES-intro-def}
	\mathrm{MES}_p(L_1,S)=\mathbb{E}[L_1\vert S> \rm{VaR}_p(S)].
\end{equation*}\cite{chen2025marginal} derived the lower and upper bounds for $\mathrm{MES}_p(L_1, S)$ when the marginal distributions are known, but the dependence structure is either completely unspecified or only partially known.

As in the previous subsection, we model the individual and systemic risks by $L_1=-\frac{\ln(1-U_{1,\lambda})}{\lambda_1}$, and $S=-\frac{\ln(1-U_{1,\lambda})}{\lambda_1}-\frac{\ln(1-U_2)}{\lambda_2},$ where $U_{1,\lambda}$ and $U_2$ are as defined in Equations~\eqref{Exp copula for cosk}. Under this setup, the MES is expressed as
\begin{equation}\label{MES formula}
	\begin{aligned}
		&\mathrm{MES}_p(L_1,S)=\frac{1}{\lambda_1}\\
		&\mathbb{E}[-\ln(1-U_{1,\lambda})\vert -\ln((1-U_{1,\lambda})^{\lambda_2})(1-U_2)^{\lambda_1}> \rm{VaR}_p(-\ln((1-U_{1,\lambda})^{\lambda_2}(1-U_2)^{\lambda_1}))],
	\end{aligned}
\end{equation} where $U_{1,\lambda}$ and $U_2$ are as in \eqref{Exp copula for cosk}.
\begin{figure}
	\centering
	\begin{tikzpicture}
	\begin{axis}[
		xlabel={$p$},
		ylabel={$\mathrm{MES}_p(L_1,S)$},
		xmin=0.74, xmax=1.01,
		ymin=1.3, ymax=5,
		axis lines=middle,
		width=0.75\textwidth,
		height=0.5\textwidth,
		xlabel style={at={(current axis.south)}, below=5mm},
		ylabel style={at={(current axis.west)}, rotate=90, yshift=10mm},
		xtick={0.75, 0.8, 0.85, 0.9, 0.95, 1},
		legend style={at={(0.15,0.85)}, anchor=north}
		]
		\addplot[] coordinates {
			(0.75, 1.4909778849549955)
			(0.7525252525252525, 1.4969648408795193)
			(0.7550505050505051, 1.5033987682163952)
			(0.7575757575757576, 1.5097918069060838)
			(0.76010101010101, 1.516237906105163)
			(0.7626262626262627, 1.5228063222614074)
			(0.7651515151515151, 1.5296248410126152)
			(0.7676767676767677, 1.5363677988219377)
			(0.7702020202020202, 1.5431351553033585)
			(0.7727272727272727, 1.5500314347784652)
			(0.7752525252525253, 1.5567414860806752)
			(0.7777777777777778, 1.5636219182698923)
			(0.7803030303030303, 1.5709621690591602)
			(0.7828282828282829, 1.5781886438065265)
			(0.7853535353535354, 1.5851117157252541)
			(0.7878787878787878, 1.5921679222084104)
			(0.7904040404040404, 1.5994408272720586)
			(0.7929292929292929, 1.6066258013791332)
			(0.7954545454545454, 1.6138858831989185)
			(0.797979797979798, 1.621538495501687)
			(0.8005050505050505, 1.629272679445997)
			(0.803030303030303, 1.6369151921803486)
			(0.8055555555555556, 1.6450657422355883)
			(0.8080808080808081, 1.6534707010813474)
			(0.8106060606060606, 1.663864949710379)
			(0.8131313131313131, 1.6794547143616751)
			(0.8156565656565656, 1.695149930124825)
			(0.8181818181818182, 1.7110646986004374)
			(0.8207070707070707, 1.7270670408357722)
			(0.8232323232323232, 1.7432888364043642)
			(0.8257575757575758, 1.7595986130966623)
			(0.8282828282828283, 1.7761034868906806)
			(0.8308080808080808, 1.792628401063363)
			(0.8333333333333334, 1.8093020222997838)
			(0.8358585858585859, 1.8260258597155765)
			(0.8383838383838385, 1.8426728652453865)
			(0.8409090909090909, 1.8594172325958451)
			(0.8434343434343434, 1.8760477545608172)
			(0.845959595959596, 1.8925888607999908)
			(0.8484848484848485, 1.9088594821239058)
			(0.851010101010101, 1.9249789045033907)
			(0.8535353535353536, 1.9406966721627192)
			(0.8560606060606061, 1.9559989049132638)
			(0.8585858585858586, 1.9706707531007728)
			(0.8611111111111112, 1.9845099641307613)
			(0.8636363636363636, 1.9971427313038728)
			(0.8661616161616161, 2.0095342402951477)
			(0.8686868686868687, 2.022121760638265)
			(0.8712121212121212, 2.03501272197332)
			(0.8737373737373737, 2.0481021385147935)
			(0.8762626262626263, 2.061473463320988)
			(0.8787878787878788, 2.075069972538959)
			(0.8813131313131313, 2.0890072308636927)
			(0.8838383838383839, 2.1032337981475555)
			(0.8863636363636364, 2.117845857598284)
			(0.8888888888888888, 2.1327899890590833)
			(0.8914141414141414, 2.1480279973983745)
			(0.8939393939393939, 2.1636781780606986)
			(0.8964646464646464, 2.1796596797015257)
			(0.898989898989899, 2.1960796821608497)
			(0.9015151515151515, 2.21283477746698)
			(0.9040404040404041, 2.230124885997997)
			(0.9065656565656566, 2.2478368870662204)
			(0.9090909090909092, 2.2661185358637512)
			(0.9116161616161617, 2.2848636535523825)
			(0.9141414141414141, 2.3042483628014647)
			(0.9166666666666667, 2.3241400893392297)
			(0.9191919191919192, 2.344702053473115)
			(0.9217171717171717, 2.3658299103559304)
			(0.9242424242424243, 2.387722014858566)
			(0.9267676767676768, 2.4103022598351185)
			(0.9292929292929293, 2.433833636687936)
			(0.9318181818181819, 2.458121732193508)
			(0.9343434343434344, 2.483462859008038)
			(0.9368686868686869, 2.5096497762526875)
			(0.9393939393939394, 2.5370108106416795)
			(0.9419191919191919, 2.5654260501422503)
			(0.9444444444444444, 2.5953225883401694)
			(0.946969696969697, 2.6266492032439683)
			(0.9494949494949495, 2.6593660543098836)
			(0.952020202020202, 2.6938297910807294)
			(0.9545454545454546, 2.72997814038257)
			(0.9570707070707071, 2.7683859791469074)
			(0.9595959595959596, 2.809036045213076)
			(0.9621212121212122, 2.8527399513316825)
			(0.9646464646464646, 2.8995030195185296)
			(0.9671717171717171, 2.9501455187142023)
			(0.9696969696969697, 3.0048229228746584)
			(0.9722222222222222, 3.0642785248569417)
			(0.9747474747474748, 3.128739845354276)
			(0.9772727272727273, 3.199441920079579)
			(0.9797979797979798, 3.2776148698558574)
			(0.9823232323232324, 3.3657913371884662)
			(0.9848484848484849, 3.467232827768161)
			(0.9873737373737375, 3.587480487695327)
			(0.98989898989899, 3.7365710312291567)
			(0.9924242424242424, 3.9355401498646803)
			(0.994949494949495, 4.224416840115107)
			(0.9974747474747475, 4.7106401837752765)
		};
		\addlegendentry{$\lambda=1$}
		\addplot[dashed] coordinates {(0.75, 1.3949731852599525)
			(0.7525252525252525, 1.4010855365084183)
			(0.7550505050505051, 1.407767443405718)
			(0.7575757575757576, 1.4140689708571157)
			(0.76010101010101, 1.4203671976228058)
			(0.7626262626262627, 1.4271995321189694)
			(0.7651515151515151, 1.4336355658495858)
			(0.7676767676767677, 1.4406312646613924)
			(0.7702020202020202, 1.4473250177684818)
			(0.7727272727272727, 1.454775276956846)
			(0.7752525252525253, 1.4625541641886932)
			(0.7777777777777778, 1.4696412649317907)
			(0.7803030303030303, 1.4767621087595946)
			(0.7828282828282829, 1.484216321420976)
			(0.7853535353535354, 1.4914558932612347)
			(0.7878787878787878, 1.498552931111526)
			(0.7904040404040404, 1.5066637331451134)
			(0.7929292929292929, 1.5146086890823356)
			(0.7954545454545454, 1.5217677070349636)
			(0.797979797979798, 1.529623290339975)
			(0.8005050505050505, 1.537755640540701)
			(0.803030303030303, 1.5459634860173916)
			(0.8055555555555556, 1.5542253029498625)
			(0.8080808080808081, 1.5618793797488164)
			(0.8106060606060606, 1.5702839722326922)
			(0.8131313131313131, 1.5790232548277334)
			(0.8156565656565656, 1.5870084680256198)
			(0.8181818181818182, 1.5961655014345157)
			(0.8207070707070707, 1.6038408373642867)
			(0.8232323232323232, 1.6121164641580827)
			(0.8257575757575758, 1.6200846481059883)
			(0.8282828282828283, 1.6292253662930127)
			(0.8308080808080808, 1.6380907861152376)
			(0.8333333333333334, 1.64823902049069)
			(0.8358585858585859, 1.657138729343857)
			(0.8383838383838385, 1.6666137881758945)
			(0.8409090909090909, 1.6754760290798907)
			(0.8434343434343434, 1.6854018342861485)
			(0.845959595959596, 1.6956906209975011)
			(0.8484848484848485, 1.7053127130883317)
			(0.851010101010101, 1.7151570334712491)
			(0.8535353535353536, 1.7256782425007782)
			(0.8560606060606061, 1.7365490211677903)
			(0.8585858585858586, 1.747298897940273)
			(0.8611111111111112, 1.7591713272828358)
			(0.8636363636363636, 1.7702355559613778)
			(0.8661616161616161, 1.7809922216015968)
			(0.8686868686868687, 1.7926528175051433)
			(0.8712121212121212, 1.804894672528289)
			(0.8737373737373737, 1.8172448572944522)
			(0.8762626262626263, 1.8287692479633184)
			(0.8787878787878788, 1.8417160087907967)
			(0.8813131313131313, 1.8553791812609368)
			(0.8838383838383839, 1.8695423392107464)
			(0.8863636363636364, 1.8842768284082478)
			(0.8888888888888888, 1.8975984046956809)
			(0.8914141414141414, 1.9121216180179745)
			(0.8939393939393939, 1.9269090766272545)
			(0.8964646464646464, 1.94144113809701)
			(0.898989898989899, 1.955600831182833)
			(0.9015151515151515, 1.9716249021926882)
			(0.9040404040404041, 1.9892424661455215)
			(0.9065656565656566, 2.006651476139529)
			(0.9090909090909092, 2.025082285808644)
			(0.9116161616161617, 2.0434775399785283)
			(0.9141414141414141, 2.062397646876174)
			(0.9166666666666667, 2.0840705390831147)
			(0.9191919191919192, 2.103485785930515)
			(0.9217171717171717, 2.124715709038046)
			(0.9242424242424243, 2.145697376972047)
			(0.9267676767676768, 2.1692163086624654)
			(0.9292929292929293, 2.1954350979957353)
			(0.9318181818181819, 2.2190404419607943)
			(0.9343434343434344, 2.243452306700827)
			(0.9368686868686869, 2.2698725101742347)
			(0.9393939393939394, 2.296589855999978)
			(0.9419191919191919, 2.3244399393812785)
			(0.9444444444444444, 2.3537477275853202)
			(0.946969696969697, 2.3839100367745174)
			(0.9494949494949495, 2.4139073281027876)
			(0.952020202020202, 2.4434412536425714)
			(0.9545454545454546, 2.4781017514614105)
			(0.9570707070707071, 2.5127401709376205)
			(0.9595959595959596, 2.5558713200061907)
			(0.9621212121212122, 2.598709681086712)
			(0.9646464646464646, 2.6422385398481505)
			(0.9671717171717171, 2.6879481391888014)
			(0.9696969696969697, 2.7404553092885027)
			(0.9722222222222222, 2.792042387752213)
			(0.9747474747474748, 2.8504197621265126)
			(0.9772727272727273, 2.9184950438454482)
			(0.9797979797979798, 3.0020237235918747)
			(0.9823232323232324, 3.080152618325842)
			(0.9848484848484849, 3.1688531002382025)
			(0.9873737373737375, 3.303522369014978)
			(0.98989898989899, 3.465573145066217)
			(0.9924242424242424, 3.6640027266828814)
			(0.994949494949495, 3.9626733032837524)
			(0.9974747474747475, 4.459870293684094)
		};
		\addlegendentry{$\lambda=0$}
		\addplot [] coordinates {(0,4.9) (1,4.9)};
		\addplot [] coordinates {(1,0) (1,4.9)};
	\end{axis}
\end{tikzpicture}
	\caption{$\mathrm{MES}_p(L_1,S)$ with $\overline{S_{2k}}$ and $\underline{S_{2k}}$, where $L_1\sim \mathrm{Expon}(1.5)$ and $L_2\sim \mathrm{Expon}(2)$, as functions of the prudence level $p\in[0.75, 1)$.}
	\label{range of MES}
\end{figure}
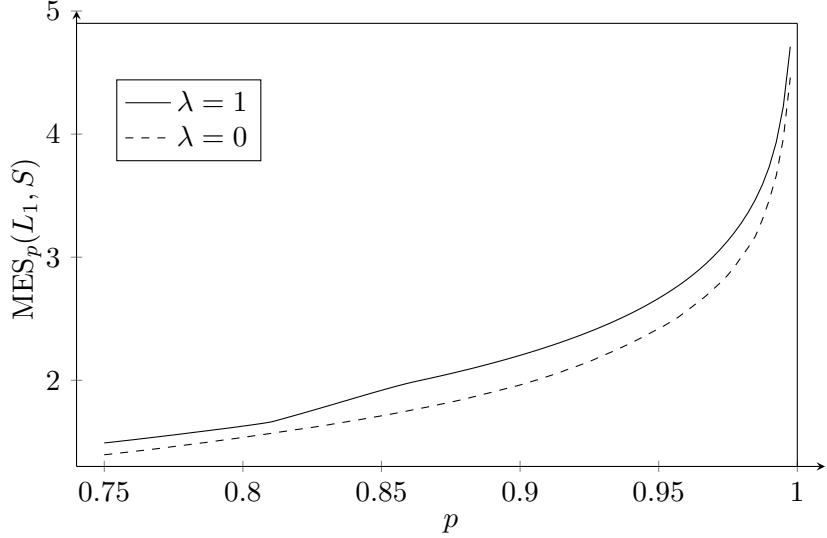

As with the ES, the MES is computed numerically using Equation~\eqref{MES formula} with $n = 10^6$ independent samples drawn from the standard uniform distribution. In Figure~\ref{range of MES}, we present $\mathrm{MES}_p(L_1, S)$ under the minimum and maximum values of $\mu_{2k}(X_1,X_2)$, plotted against the prudence level $p \in [0.75, 1)$ for fixed parameters $\lambda_1 = 1.5$ and $\lambda_2 = 2$. Consistent with the behavior of ES, MES under the maximum mixed moment ($\lambda = 1$) is constantly greater than under the minimum ($\lambda = 0$). However, unlike ES, the difference between the two MES values initially widens with increasing $p$, but then narrows, eventually vanishing as $p \to 1$. This non-monotonic trend is further illustrated in Table~\ref{effect of cosk on mes}, which reports MES values for selected combinations of $\lambda_1$, $\lambda_2$, and $p$. In Panels A, B, D, and E, MES increases monotonically with $\lambda$, indicating a positive correlation between MES and the mixed moment $\mu_{2k}(X_1,X_2)$. However, Panel C exhibits a different behavior: MES remains largely unchanged as $\mu_{2k}(X_1,X_2)$ increases. For instance, the MES values at $\lambda = 0$ and $\lambda = 0.5$ are equivalent, suggesting a lack of sensitivity to $\mu_{2k}(X_1,X_2)$ in that parameter configuration.
 
\begin{table}[!h]
	\centering 
	\caption{The effect of $\mu_{2k}(X_1,X_2)$ on $\mathrm{MES}_p(L_1,S)$, where $L_1\sim \mathrm{Expon}(\lambda_1)$ and $L_2\sim \mathrm{Expon}(\lambda_2)$.}
	\begin{tabularx}{\textwidth}{*{6}{X}}
		\toprule
		$\lambda$& 0 & 0.25 & 0.5 & 0.75 & 1.00\\
		\midrule
		\multicolumn{6}{l}{Panel A: benchmark case ($\lambda_1=\lambda_2=1$ and $p=0.95$)} \\
		$\mathrm{MES}_p(L_1,S)$ & 3.18 & 3.35 & 3.64 & 3.79 & 4.02 \\
		\midrule
		\multicolumn{6}{l}{Panel B: changing rate parameter for $L_1$ ($\lambda_1=2,$ $\lambda_2=1$ and $p=0.95$)}\\
		$\mathrm{MES}_p(L_1,S)$ & 0.62 & 1.05 & 1.47 & 1.78 & 1.99 \\
		\midrule
		\multicolumn{6}{l}{Panel C: changing rate parameter for $L_2$ ($\lambda_1=1,$ $\lambda_2=2$ and $p=0.95$)}\\
		$\mathrm{MES}_p(L_1,S)$ & 3.91 & 3.88 & 3.91 & 3.92 & 4.00   \\
		\midrule
		\multicolumn{6}{l}{Panel D: reducing the prudence level ($\lambda_1=\lambda_2=1$ and $p=0.90$)}\\
		$\mathrm{MES}_p(L_1,S)$ & 2.65 & 2.8 & 2.99 & 3.15 & 3.33 \\
		\midrule
		\multicolumn{6}{l}{Panel E: rising the prudence level ($\lambda_1=\lambda_2=1$ and $p=0.99$)}\\
		$\mathrm{MES}_p(L_1,S)$ & 4.19 & 4.6 & 4.97 & 5.24 & 5.53 \\
		\bottomrule
	\end{tabularx} 
	\label{effect of cosk on mes}
\end{table}

The findings of this subsection indicate that the influence of $(2k-1)$-th-order standardized centered mixed moments, such as coskewness, on MES is nuanced and context-dependent. While in many cases MES increases with coskewness, this is not a universal rule. Therefore, caution must be exercised when interpreting the role of mixed moments in systemic risk contribution. Importantly, risk managers may not be able to reduce systemic exposure simply by selecting individual risks with lower coskewness, as the MES may remain unchanged or even increase depending on the specific dependence structure and distributional parameters.

\subsection{Life annuities}\label{pricing life annuit}
In this subsection, we study how the $(2k-1)$-th-order standardized centered mixed moments $\mu_{2k}(X_1,X_2)$, serving as a measure of dependence, affects joint-life and last-survivor probabilities for insured couples, and consequently, the net single premiums of corresponding life annuity products. 

We begin by introducing the notation for life contingent risks used through this subsection. We refer readers to \cite{denuit2001measuring} and \cite{dickson2020actuarial} for basic life insurance theory. Let $T_x \sim \mathrm{Expon}(\lambda_1)$ and $T_y \sim \mathrm{Expon}(\lambda_2)$ denote the future lifetimes of a married male aged $x$ and his wife aged $y$, respectively. Without loss of generality, we only focus on studying the impact of the mixed moments $\mu_{2k}(X_1,X_2)=\mathbb{E}\left[\frac{T_x-\mu_1}{\sigma_1}\left(\frac{T_y-\mu_2}{\sigma_2}\right)^{2k}\right]$, while omitting the symmetric case (with roles of $T_x$ and $T_y$ reversed), as the findings are qualitatively similar. 

Let ${}_tq_x$ and ${}_tq_y$ (resp.\ ${}_tp_x$ and ${}_tp_y$) denote the CDFs (resp.\ survival functions) of $T_x$ and $T_y$, respectively. Specifically,
\begin{equation*}
	{}_tp_x = \mathbb{P}(T_x>t)=e^{-\lambda_1 t}=1-{}_tq_x,\ t\in \mathbb{R}^+,
\end{equation*}and
\begin{equation*}
	{}_tp_y = \mathbb{P}(T_y>t)=e^{-\lambda_2 t}=1-{}_tq_y,\ t\in \mathbb{R}^+.
\end{equation*} The probability of the joint-life status $\min(T_x, T_y)$ surviving to time $t$, denoted by ${}_tp_{x:y}$, is given by
\begin{equation*}
	\begin{aligned}
		{}_tp_{x:y,\lambda}&=\mathbb{P}(\min(T_x, T_y)>t)=\mathbb{P}(T_x>t, T_y>t)\\
		&=\mathbb{P}(U< e^{-\lambda_1 t}, U_2> 1-e^{-\lambda_2 t})(1-\lambda)+\mathbb{P}(U> 1-e^{-\lambda_1 t}, U_2> 1-e^{-\lambda_2 t})\lambda,
	\end{aligned}
\end{equation*} where $U\sim U[0,1]$ and $U_2$ is as defined in Equations~\eqref{Exp copula for cosk}. Similarly, the probability that the last-survivor status $\max(T_x, T_y)$ surviving to time $t$, denoted by ${}_tp_{\overline{x:y}}$, is given by
\begin{equation*}
	\begin{aligned}
		{}_tp_{\overline{x:y}, \lambda}=&\mathbb{P}(\max(T_x, T_y)>t)=\mathbb{P}(T_x>t \cup T_y>t)={}_tp_x+{}_tp_y-{}_tp^{\lambda}_{x:y}\\
		=&e^{-\lambda_1 t}+e^{-\lambda_2 t}-\mathbb{P}(U< e^{-\lambda_1 t}, U_2> 1-e^{-\lambda_2 t})(1-\lambda)\\
		&-\mathbb{P}(U> 1-e^{-\lambda_1 t}, U_2> 1-e^{-\lambda_2 t})\lambda.
	\end{aligned}
\end{equation*} 

Let $\nu = (1 + \xi)^{-1}$ denote the discount factor associated with a constant annual effective interest rate $\xi$. In this study, we set $\xi = 4.5\%$, consistent with prior literature; see \cite{denuit2001measuring}, \cite{hsieh2021mortality}, and \cite{deresa2022copula}.

We hereafter consider the joint-life and last-survivor annuities, which are contractual guarantees that promise to offer periodic income over the lifetimes of dependent people \citep[see][]{denuit2001measuring, dickson2020actuarial}. In the case of a married couple, the $n$-year last-survivor and joint-life annuities pay 1 dollar at the end of years $1, 2, \dots, n$, as long as either spouse survives and both spouses survive, are defined, respectively, as
\begin{equation*}
	a_{\overline{x:y};\overline{n}\vert,\lambda}=\sum_{k=1}^{n}\nu^k {}_kp_{\overline{x:y},\lambda},
\end{equation*}and
\begin{equation*}
	a_{x:y;\overline{n}\vert,\lambda}=\sum_{k=1}^{n}\nu^k {}_kp_{x:y,\lambda}.
\end{equation*}
For comparison, the corresponding annuity values under the assumption of independence are
\begin{equation*}
	a_{\overline{x:y};\overline{n}\vert,\perp}=\sum_{k=1}^{n}\nu^k {}_kp_{\overline{x:y},\perp}=\sum_{k=1}^{n}\nu^k ({}_kp_{x}+{}_kp_{y}-{}_kp_{x}{}_kp_{y})=\sum_{k=1}^{n}\nu^k(e^{-\lambda_1 k}+e^{-\lambda_2 k}-e^{-\lambda_1k-\lambda_2k}),
\end{equation*}and
\begin{equation*}
	a_{x:y;\overline{n}\vert,\perp}=\sum_{k=1}^{n}\nu^k {}_kp_{x:y,\perp}=\sum_{k=1}^{n}\nu^k {}_kp_{x}{}_kp_{y}=\sum_{k=1}^{n}\nu^k e^{-\lambda_1k-\lambda_2k}.
\end{equation*}The assumption of independence is commonly adopted in traditional actuarial models of multiple-life insurance products for analytical tractability; see \cite{denuit2001measuring}.

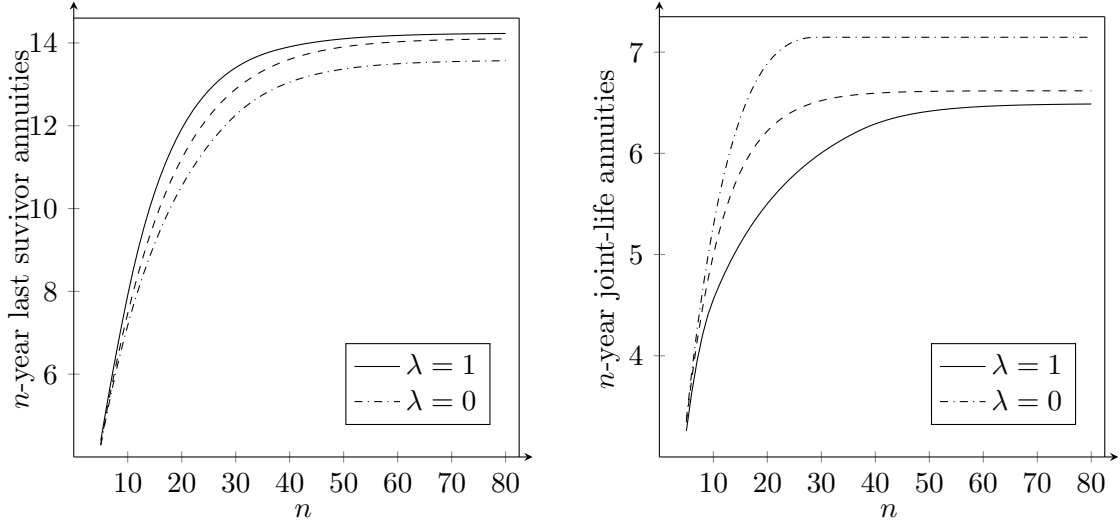
\begin{figure}
	\begin{subfigure}{0.5\textwidth}
		\centering
		\begin{tikzpicture}
\begin{axis}[
	xlabel={$n$},
	ylabel={$n$-year last suvivor annuities},
	xmin=0, xmax=85,
	ymin=4, ymax=15,
	axis lines=middle,
	width=\textwidth,
	height=\textwidth,
	xlabel style={at={(current axis.south)}, below=5mm},
	ylabel style={at={(current axis.west)}, rotate=90, yshift=10mm},
	xtick={10, 20, 30, 40, 50, 60, 70, 80},
	legend style={at={(0.75,0.25)}, anchor=north}
	]
	\addplot[] coordinates {
		(5.0, 4.389531013737203)
		(6.0, 5.157102322401243)
		(7.0, 5.891221540664658)
		(8.0, 6.594019303447327)
		(9.0, 7.266486220075882)
		(10.0, 7.909469940396708)
		(11.0, 8.506077513231334)
		(12.0, 9.048703010007628)
		(13.0, 9.54163301008707)
		(14.0, 9.989535875897225)
		(15.0, 10.396023145754542)
		(16.0, 10.764869484507479)
		(17.0, 11.099273577506613)
		(18.0, 11.40240030656986)
		(19.0, 11.677047447584659)
		(20.0, 11.925648062687635)
		(21.0, 12.150507240749675)
		(22.0, 12.353757384811228)
		(23.0, 12.537389329197053)
		(24.0, 12.703101128042505)
		(25.0, 12.852599873179711)
		(26.0, 12.987326836112183)
		(27.0, 13.108593963872332)
		(28.0, 13.217709364269194)
		(29.0, 13.315828908017577)
		(30.0, 13.403948468970098)
		(31.0, 13.48304798093219)
		(32.0, 13.553920615136924)
		(33.0, 13.61728639100968)
		(34.0, 13.673819206845321)
		(35.0, 13.724211742165954)
		(36.0, 13.769093505422815)
		(37.0, 13.809001054775528)
		(38.0, 13.845012465216607)
		(39.0, 13.877934931263951)
		(40.0, 13.90805920286094)
		(41.0, 13.935647319201719)
		(42.0, 13.960933856846124)
		(43.0, 13.984088073875405)
		(44.0, 14.005285458544593)
		(45.0, 14.024715259090483)
		(46.0, 14.042508274660873)
		(47.0, 14.058825112200537)
		(48.0, 14.073793408942594)
		(49.0, 14.087549331458082)
		(50.0, 14.100189823890087)
		(51.0, 14.11178680660383)
		(52.0, 14.122399965930766)
		(53.0, 14.13212211610206)
		(54.0, 14.141023068269606)
		(55.0, 14.149169788037923)
		(56.0, 14.156642577101424)
		(57.0, 14.163481882863932)
		(58.0, 14.169749809296338)
		(59.0, 14.175501590935053)
		(60.0, 14.180793437271884)
		(61.0, 14.185656710609756)
		(62.0, 14.190109596322422)
		(63.0, 14.194196043477325)
		(64.0, 14.197934646981613)
		(65.0, 14.20138330774149)
		(66.0, 14.20454260602091)
		(67.0, 14.207433252748752)
		(68.0, 14.210080686436848)
		(69.0, 14.21250213650997)
		(70.0, 14.21472116006604)
		(71.0, 14.216755462711797)
		(72.0, 14.218618563471116)
		(73.0, 14.220318377869663)
		(74.0, 14.221874649263878)
		(75.0, 14.223301411382389)
		(76.0, 14.224610980957767)
		(77.0, 14.225812123023406)
		(78.0, 14.226912830014797)
		(79.0, 14.227918544469267)
		(80.0, 14.228840527795015)
	};
	\addlegendentry{$\lambda=1$}
	\addplot[dash dot] coordinates {
		(5.0, 4.277993350876271)
		(6.0, 4.954745630574152)
		(7.0, 5.578934511568272)
		(8.0, 6.154905804850838)
		(9.0, 6.6873682262476795)
		(10.0, 7.179608476205271)
		(11.0, 7.63560565883661)
		(12.0, 8.058387871717585)
		(13.0, 8.451549002948303)
		(14.0, 8.816479638749737)
		(15.0, 9.156036109713892)
		(16.0, 9.472839552243158)
		(17.0, 9.768164007568377)
		(18.0, 10.043579354058183)
		(19.0, 10.300799097144726)
		(20.0, 10.541388812198596)
		(21.0, 10.766370994362552)
		(22.0, 10.977078463818858)
		(23.0, 11.174935565787809)
		(24.0, 11.36051166407594)
		(25.0, 11.53472896524035)
		(26.0, 11.69839553002168)
		(27.0, 11.852346901371162)
		(28.0, 11.997383811014297)
		(29.0, 12.134035328500566)
		(30.0, 12.260616241686872)
		(31.0, 12.376114626576358)
		(32.0, 12.48150331324246)
		(33.0, 12.577669512606516)
		(34.0, 12.665422371586796)
		(35.0, 12.745499859827033)
		(36.0, 12.81857504729449)
		(37.0, 12.88526182676499)
		(38.0, 12.946120130409861)
		(39.0, 13.001660685325195)
		(40.0, 13.05234934885925)
		(41.0, 13.098611060964068)
		(42.0, 13.140833447491028)
		(43.0, 13.179370105338052)
		(44.0, 13.214543597612336)
		(45.0, 13.246648184472852)
		(46.0, 13.275952313039632)
		(47.0, 13.302700887682194)
		(48.0, 13.327117340109368)
		(49.0, 13.349405516960646)
		(50.0, 13.36975140103026)
		(51.0, 13.38832468082557)
		(52.0, 13.405280181858831)
		(53.0, 13.420759171884589)
		(54.0, 13.434890551213435)
		(55.0, 13.44779193824753)
		(56.0, 13.459570659485232)
		(57.0, 13.47032465242397)
		(58.0, 13.48014328904478)
		(59.0, 13.489108126882364)
		(60.0, 13.497293594065312)
		(61.0, 13.504767614146633)
		(62.0, 13.511592176030486)
		(63.0, 13.51782385383212)
		(64.0, 13.523514281080832)
		(65.0, 13.52871058328629)
		(66.0, 13.53345577253364)
		(67.0, 13.53778910744929)
		(68.0, 13.541746421584355)
		(69.0, 13.545360422993971)
		(70.0, 13.548660967545619)
		(71.0, 13.551675308266264)
		(72.0, 13.554428322834498)
		(73.0, 13.556942721138252)
		(74.0, 13.559239234649409)
		(75.0, 13.561336789212415)
		(76.0, 13.563252662703322)
		(77.0, 13.565002628887447)
		(78.0, 13.566601088686983)
		(79.0, 13.568061189963256)
		(80.0, 13.569394936821132)
		
	};
	\addlegendentry{$\lambda=0$}
	\addplot[dashed] coordinates {
	(5.0, 4.3020810462988095)
	(6.0, 5.021792179487839)
	(7.0, 5.69666957022026)
	(8.0, 6.328249293239329)
	(9.0, 6.918247813800297)
	(10.0, 7.4685046634028)
	(11.0, 7.980935611150659)
	(12.0, 8.457494690003537)
	(13.0, 8.900143675902722)
	(14.0, 9.310827823058275)
	(15.0, 9.691456835148205)
	(16.0, 10.043890203756746)
	(17.0, 10.369926175488873)
	(18.0, 10.671293720799413)
	(19.0, 10.949646973224644)
	(20.0, 11.206561689611986)
	(21.0, 11.443533352018822)
	(22.0, 11.661976591843993)
	(23.0, 11.86322566788952)
	(24.0, 12.048535773654946)
	(25.0, 12.219084986303626)
	(26.0, 12.375976701324275)
	(27.0, 12.520242423731593)
	(28.0, 12.652844809386856)
	(29.0, 12.774680869259088)
	(30.0, 12.886585265694542)
	(31.0, 12.98933364345094)
	(32.0, 13.083645949756942)
	(33.0, 13.170189707297949)
	(34.0, 13.249583212082385)
	(35.0, 13.322398634845209)
	(36.0, 13.389165010200784)
	(37.0, 13.450371102339728)
	(38.0, 13.506468139823298)
	(39.0, 13.55787241509206)
	(40.0, 13.604967746782224)
	(41.0, 13.648107804926575)
	(42.0, 13.6876183006869)
	(43.0, 13.723799043489112)
	(44.0, 13.756925869368624)
	(45.0, 13.78725244503102)
	(46.0, 13.81501195263333)
	(47.0, 13.840418660629386)
	(48.0, 13.863669386228793)
	(49.0, 13.884944855117986)
	(50.0, 13.904410964104759)
	(51.0, 13.92221995229245)
	(52.0, 13.938511486281328)
	(53.0, 13.953413664745069)
	(54.0, 13.967043947549787)
	(55.0, 13.979510014380493)
	(56.0, 13.990910557621834)
	(57.0, 14.001336014012495)
	(58.0, 14.010869239360138)
	(59.0, 14.019586130370115)
	(60.0, 14.027556197409124)
	(61.0, 14.034843091797184)
	(62.0, 14.041505090999141)
	(63.0, 14.047595544872175)
	(64.0, 14.053163285919005)
	(65.0, 14.058253006298838)
	(66.0, 14.06290560415955)
	(67.0, 14.06715850167577)
	(68.0, 14.071045937008265)
	(69.0, 14.074599232240363)
	(70.0, 14.077847039197012)
	(71.0, 14.080815564911015)
	(72.0, 14.083528778368986)
	(73.0, 14.08600860004612)
	(74.0, 14.08827507562365)
	(75.0, 14.090346535175525)
	(76.0, 14.092239739010973)
	(77.0, 14.093970011266782)
	(78.0, 14.095551362257005)
	(79.0, 14.096996600507907)
	(80.0, 14.098317435332042)
};
	\addplot [] coordinates {(0,14.6) (82.5,14.6)};
	\addplot [] coordinates {(82.5,0) (82.5,14.6)};
\end{axis}
\end{tikzpicture}	
	\end{subfigure}
	\begin{subfigure}{0.5\textwidth}
		\centering
		\begin{tikzpicture}
	\begin{axis}[
		xlabel={$n$},
		ylabel={$n$-year joint-life annuities},
		xmin=0, xmax=85,
		ymin=3, ymax=7.5,
		axis lines=middle,
		width=\textwidth,
		height=\textwidth,
		xlabel style={at={(current axis.south)}, below=5mm},
		ylabel style={at={(current axis.west)}, rotate=90, yshift=10mm},
		xtick={10, 20, 30, 40, 50, 60, 70, 80},
		legend style={at={(0.75,0.25)}, anchor=north}
		]
		\addplot[] coordinates {
			(5.0, 3.2560033752392297)
			(6.0, 3.638000789203087)
			(7.0, 3.952191392854375)
			(8.0, 4.205387261522051)
			(9.0, 4.4047486563230684)
			(10.0, 4.556857253372205)
			(11.0, 4.685377824320536)
			(12.0, 4.804088953595313)
			(13.0, 4.914330704101366)
			(14.0, 5.016563766105475)
			(15.0, 5.1118521828729255)
			(16.0, 5.200683596712187)
			(17.0, 5.283744979413233)
			(18.0, 5.361413826681568)
			(19.0, 5.43412186677012)
			(20.0, 5.502380374331397)
			(21.0, 5.566568676160273)
			(22.0, 5.627001869125085)
			(23.0, 5.6839206669640845)
			(24.0, 5.7376617158326395)
			(25.0, 5.788373186072323)
			(26.0, 5.836305496143036)
			(27.0, 5.881689276161971)
			(28.0, 5.924623060549258)
			(29.0, 5.965233696157869)
			(30.0, 6.003695048391653)
			(31.0, 6.040093921319047)
			(32.0, 6.074609973780414)
			(33.0, 6.107410397271715)
			(34.0, 6.138630440416353)
			(35.0, 6.168315393335958)
			(36.0, 6.196508817546553)
			(37.0, 6.223288047664338)
			(38.0, 6.2481349408681295)
			(39.0, 6.2707530297361185)
			(40.0, 6.291317421673184)
			(41.0, 6.309991017437223)
			(42.0, 6.326926866319779)
			(43.0, 6.342309307137521)
			(44.0, 6.356285414742617)
			(45.0, 6.368960201057243)
			(46.0, 6.380471314053632)
			(47.0, 6.390903051156531)
			(48.0, 6.400351206841648)
			(49.0, 6.408883461177439)
			(50.0, 6.416588852815048)
			(51.0, 6.423565149896615)
			(52.0, 6.4299074916029415)
			(53.0, 6.435664331457404)
			(54.0, 6.440894758618704)
			(55.0, 6.445649425884481)
			(56.0, 6.449955358058682)
			(57.0, 6.453870045234914)
			(58.0, 6.457420755423317)
			(59.0, 6.460633811622187)
			(60.0, 6.463527432468302)
			(61.0, 6.466138179211751)
			(62.0, 6.468509855382937)
			(63.0, 6.470655086029668)
			(64.0, 6.472606909774091)
			(65.0, 6.474354551219672)
			(66.0, 6.4759404421876035)
			(67.0, 6.4773831303754115)
			(68.0, 6.478693010822379)
			(69.0, 6.479885562158874)
			(70.0, 6.480967083154451)
			(71.0, 6.4819471212293385)
			(72.0, 6.482837035038254)
			(73.0, 6.483651618943461)
			(74.0, 6.484391861060403)
			(75.0, 6.485062653504899)
			(76.0, 6.485668957420427)
			(77.0, 6.486217781538911)
			(78.0, 6.4867155343470575)
			(79.0, 6.487169921168859)
			(80.0, 6.487581684700988)
		};
		\addlegendentry{$\lambda=1$}
		\addplot[dash dot] coordinates {
			(5.0, 3.3675410381001627)
			(6.0, 3.8403574810301784)
			(7.0, 4.264478421950762)
			(8.0, 4.644500760118541)
			(9.0, 4.983866650151272)
			(10.0, 5.286718717563642)
			(11.0, 5.55584967871526)
			(12.0, 5.794404091885356)
			(13.0, 6.004414711240134)
			(14.0, 6.189620003252964)
			(15.0, 6.351839218913576)
			(16.0, 6.492713528976506)
			(17.0, 6.614854549351469)
			(18.0, 6.7202347791932455)
			(19.0, 6.810370217210054)
			(20.0, 6.886639624820437)
			(21.0, 6.950704922547397)
			(22.0, 7.003680790117455)
			(23.0, 7.046374430373328)
			(24.0, 7.080251179799204)
			(25.0, 7.106244094011684)
			(26.0, 7.125236802233539)
			(27.0, 7.137936338663141)
			(28.0, 7.1449486138041545)
			(29.0, 7.147027275674881)
			(30.0, 7.147027275674881)
			(31.0, 7.147027275674881)
			(32.0, 7.147027275674881)
			(33.0, 7.147027275674881)
			(34.0, 7.147027275674881)
			(35.0, 7.147027275674881)
			(36.0, 7.147027275674881)
			(37.0, 7.147027275674881)
			(38.0, 7.147027275674881)
			(39.0, 7.147027275674881)
			(40.0, 7.147027275674881)
			(41.0, 7.147027275674881)
			(42.0, 7.147027275674881)
			(43.0, 7.147027275674881)
			(44.0, 7.147027275674881)
			(45.0, 7.147027275674881)
			(46.0, 7.147027275674881)
			(47.0, 7.147027275674881)
			(48.0, 7.147027275674881)
			(49.0, 7.147027275674881)
			(50.0, 7.147027275674881)
			(51.0, 7.147027275674881)
			(52.0, 7.147027275674881)
			(53.0, 7.147027275674881)
			(54.0, 7.147027275674881)
			(55.0, 7.147027275674881)
			(56.0, 7.147027275674881)
			(57.0, 7.147027275674881)
			(58.0, 7.147027275674881)
			(59.0, 7.147027275674881)
			(60.0, 7.147027275674881)
			(61.0, 7.147027275674881)
			(62.0, 7.147027275674881)
			(63.0, 7.147027275674881)
			(64.0, 7.147027275674881)
			(65.0, 7.147027275674881)
			(66.0, 7.147027275674881)
			(67.0, 7.147027275674881)
			(68.0, 7.147027275674881)
			(69.0, 7.147027275674881)
			(70.0, 7.147027275674881)
			(71.0, 7.147027275674881)
			(72.0, 7.147027275674881)
			(73.0, 7.147027275674881)
			(74.0, 7.147027275674881)
			(75.0, 7.147027275674881)
			(76.0, 7.147027275674881)
			(77.0, 7.147027275674881)
			(78.0, 7.147027275674881)
			(79.0, 7.147027275674881)
			(80.0, 7.147027275674881)
		};
		\addlegendentry{$\lambda=0$}
		\addplot[dashed] coordinates {
		(5.0, 3.343453342677624)
		(6.0, 3.77331093211649)
		(7.0, 4.146743363298772)
		(8.0, 4.47115727173005)
		(9.0, 4.752987062598654)
		(10.0, 4.997822530366114)
		(11.0, 5.210519726401213)
		(12.0, 5.395297273599406)
		(13.0, 5.555820038285717)
		(14.0, 5.695271818944427)
		(15.0, 5.816418493479265)
		(16.0, 5.921662877462922)
		(17.0, 6.013092381430976)
		(18.0, 6.092520412452018)
		(19.0, 6.161522341130138)
		(20.0, 6.221466747407049)
		(21.0, 6.273542564891129)
		(22.0, 6.318782662092323)
		(23.0, 6.358084328271621)
		(24.0, 6.392227070220201)
		(25.0, 6.421888072948411)
		(26.0, 6.447655630930948)
		(27.0, 6.4700408163027125)
		(28.0, 6.489487615431598)
		(29.0, 6.506381734916362)
		(30.0, 6.5210582516672115)
		(31.0, 6.533808258800299)
		(32.0, 6.544884639160399)
		(33.0, 6.5545070809834485)
		(34.0, 6.562866435179293)
		(35.0, 6.570128500656706)
		(36.0, 6.576437312768586)
		(37.0, 6.581918000100143)
		(38.0, 6.586679266261443)
		(39.0, 6.590815545908015)
		(40.0, 6.594408877751906)
		(41.0, 6.5975305317123745)
		(42.0, 6.60024242247901)
		(43.0, 6.602598337523821)
		(44.0, 6.604645003918594)
		(45.0, 6.6064230151167145)
		(46.0, 6.607967636081184)
		(47.0, 6.609309502727693)
		(48.0, 6.61047522955546)
		(49.0, 6.611487937517546)
		(50.0, 6.612367712600388)
		(51.0, 6.6131320042080075)
		(52.0, 6.613795971252391)
		(53.0, 6.614372782814407)
		(54.0, 6.6148738793385355)
		(55.0, 6.615309199541923)
		(56.0, 6.615687377538283)
		(57.0, 6.616015914086362)
		(58.0, 6.616301325359528)
		(59.0, 6.616549272187137)
		(60.0, 6.616764672331075)
		(61.0, 6.6169517980243375)
		(62.0, 6.617114360706231)
		(63.0, 6.617255584634831)
		(64.0, 6.617378270836713)
		(65.0, 6.617484852662337)
		(66.0, 6.617577444048975)
		(67.0, 6.617657881448403)
		(68.0, 6.6177277602509745)
		(69.0, 6.617788466428494)
		(70.0, 6.617841204023493)
		(71.0, 6.6178870190301335)
		(72.0, 6.617926820140396)
		(73.0, 6.617961396767015)
		(74.0, 6.617991434700642)
		(75.0, 6.618017529711774)
		(76.0, 6.618040199367233)
		(77.0, 6.6180598932955474)
		(78.0, 6.6180770021048625)
		(79.0, 6.618091865130232)
		(80.0, 6.618104777163972)
			};
		\addplot [] coordinates {(0,7.35) (82.5,7.35)};
		\addplot [] coordinates {(82.5,0) (82.5,7.35)};
	\end{axis}
\end{tikzpicture}
	\end{subfigure}
	\caption{$a_{\overline{66:64};\overline{n}\vert,\lambda}$ (left) and $a_{66:64;\overline{n}\vert,\lambda}$ (right) with $\overline{S_{2k}}$ and $\underline{S_{2k}}$, together with $a_{\overline{x:y};\overline{n}\vert,\perp}$ (left dashed line) and $a_{x:y;\overline{n}\vert,\perp}$ (right dashed line), where $T_{66}\sim \mathrm{Expon}(0.0533)$ and $T_{64}\sim \mathrm{Expon}(0.0434)$, as functions of $n$.}
	\label{range of life annuities s122}
\end{figure}

We calibrate the parameters $\lambda_1$ and $\lambda_2$ using the Canadian life table excluding Prince Edward Island \citep{statcan_life_expectancy} and the retirement age table \citep{statcan_ave_re} for the reference year 2023. We assume that males and females begin collecting annuity payments at the average retirement ages, rounded to the nearest integers: $x = 66$ for males and $y = 64$ for females. The average remaining lifetimes at those ages are 18.76 and 23.06 years, respectively. Hence, $\lambda_1 = 1/18.76 \approx 0.0533$ and $\lambda_2 = 1/23.06 \approx 0.0434$, implying $T_{66}\sim \mathrm{Expon}(0.0533)$ and $T_{64}\sim \mathrm{Expon}(0.0434)$.

Using Monte Carlo simulations with the same numerical setup as earlier subsections, Figure~\ref{range of life annuities s122} presents the $n$-year last-survivor (left panel) and joint-life (right panel) annuities under both the minimum and maximum values of $\mu_{2k-1}(X_1,X_2)$, plotted against the expiration year $n$. Several key observations follow: First, the last-survivor annuity increases with the level of dependence measured by the mixed moment, whereas the joint-life annuity decreases. Second, the gap between the values under maximum and minimum dependence grows with the annuity term $n$. Third, the differences in annuity values between the two dependence levels are comparable in magnitude for both products. 

Table~\ref{effect of cosk on life annuit} further quantifies these effects for various choices of $n$. In all cases, the last-survivor annuity is increasing in $\mu_{2k}(X_1,X_2)$, while the joint-life annuity is decreasing. Notably, the absolute differences between the maximum and minimum values are nearly identical for both annuity types. For example, $\lvert a_{\overline{66:64};\overline{30}\vert,0}-a_{\overline{66:64};\overline{30}\vert,1}\rvert=1.16$ and $\lvert a_{66:64;\overline{30}\vert,0}-a_{66:64;\overline{30}\vert,1}\rvert=1.17$.

\begin{table}[!h]
	\centering 
	\caption{The effect of $\mu_{2k}(X_1,X_2)$ on last-survivor and joint-life annuities. The future lifetimes of a couple at their ages 66 and 64 are modeled by random variables $T_{66}\sim \mathrm{Expon}(0.0533)$ and $T_{64}\sim \mathrm{Expon}(0.0434)$, respectively.}
	\begin{tabularx}{\textwidth}{*{6}{X}}
		\toprule
		$\lambda$& 0 & 0.25 & 0.5 & 0.75 & 1.00\\
		\midrule
		\multicolumn{6}{l}{Panel A: benchmark case ($n=30$)} \\
		$a_{\overline{66:64};\overline{30}\vert,\lambda}$ & 12.24 & 12.53 & 12.82 & 13.11 & 13.40 \\
		$a_{66:64;\overline{30}\vert,\lambda}$ & 7.17 & 6.88 & 6.59 & 6.3 & 6.00 \\
		\midrule
		\multicolumn{6}{l}{Panel B: reducing the end of years ($n=10$)}\\
		$a_{\overline{66:64};\overline{10}\vert,\lambda}$ & 7.17 & 7.35 & 7.54 & 7.73 & 7.91 \\
		$a_{66:64;\overline{10}\vert,\lambda}$ & 5.30 & 5.11 & 4.93 & 4.74 & 4.55 \\
		\midrule
		\multicolumn{6}{l}{Panel C: rising the end of years ($n=50$)}\\
		$a_{\overline{66:64};\overline{50}\vert,\lambda}$ & 13.35 & 13.54 & 13.73 & 13.92 & 14.11 \\
		$a_{66:64;\overline{50}\vert,\lambda}$ & 7.17 & 6.98 & 6.79 & 6.60 & 6.41 \\
		\bottomrule
	\end{tabularx} 
	\label{effect of cosk on life annuit}
\end{table}

In conclusion, this subsection demonstrates that $(2k-1)$-th-order standardized centered mixed moments - capturing dependence between future lifetimes - can significantly influence the valuation of life annuity products involving multiple lives. Assuming independence between lifetimes may lead to material mispricing. Actuaries should exercise caution and consider appropriate dependence structures to ensure actuarially fair and financially sound premium calculations.

\section{Conclusion} \label{conclusion}

This paper investigates the role of higher-order mixed moments of the form $\mathbb{E}(X_1X_2^d)$, where $X_i \sim F_i$ for $i = 1, 2$, which are crucial for understanding risk in finance and insurance. We first derive sharp lower and upper bounds for these moments under known marginal distributions but unspecified dependence structures. Furthermore, we identify the explicit dependence structures that attain these bounds for arbitrary marginals and all positive integer orders $d$. Building on these results, we apply the analytic result to centered mixed moments and introduce novel dependence measures - standardized rank coefficients - which are bounded within $[-1, 1]$ and invariant under strictly increasing transformations. In the second part of the paper, we employ a copula-based mixture model to explore how higher-order centered mixed moments influence ES, MES, and life annuity valuation. Our results reveal that, under the proposed dependence structure, coskewness and other odd-order mixed moments have a monotonic effect on ES and annuity premiums. However, MES may remain unchanged depending on the marginal distributions and model parameters. These findings offer practical implications for actuaries, risk managers, and regulators. For instance, insurers may reduce capital requirements under regulatory constraints by introducing product lines that exhibit lower odd-order mixed moments relative to the existing portfolio. However, care must be taken - lowering coskewness at the asset level does not guarantee a reduction in systemic risk contribution, as MES may remain unaffected. In the context of life annuities, the mis-specification of dependence structures, especially regarding higher-order mixed moments between future lifetimes, can lead to misleading pricing errors.

\section*{Acknowledgments}
The authors gratefully acknowledge funding from Fonds Wetenschappelijk Onderzoek (grant numbers FWO SBOS006721N and FWO G015320N). We thank Christopher Blier-Wong for his valuable hints and comments which helped to improve the paper. Jinghui Chen would like to thank funding from the project ``New Order of Risk Management: Theory and Applications in the Era of Systemic Risk'' of Natural Sciences and Engineering Research Council of Canada (NSERC).

\bibliographystyle{chicago}
\bibliography{Bibliography}
\end{document}